\documentclass[12pt]{article}
\pdfoutput=1
\usepackage[T1]{fontenc}
\usepackage{graphicx,color}
\usepackage{amsmath,amssymb,amsfonts}
\usepackage{siunitx}
\usepackage{array}
\usepackage{rotating}
\usepackage{slashed, cancel}
\usepackage{xcolor}
\usepackage{comment}
\usepackage[caption=false]{subfig}
\usepackage{braket}
\usepackage{float}
\numberwithin{equation}{section}

\restylefloat{table}
\usepackage{tabularx}
\usepackage{booktabs}

\usepackage[margin=1.0in]{geometry}
\newcolumntype{C}{>{\centering\arraybackslash}X}
\usepackage{multirow} \usepackage{array}

\title{}
\date{}

\usepackage[numbers,sort&compress]{natbib}
\usepackage[colorlinks=true, allcolors=teal]{hyperref}

\begin{document}

\begin{titlepage}

\vspace*{-2.0truecm}

\begin{flushright}
Nikhef-2025-019 
\end{flushright}

\vspace*{1.3truecm}

\begin{center}
{
\Large \bf \boldmath Standard Model Benchmarks for\\[3mm] $D^0\to K^- K^+, \pi^-\pi^+, K^0_{\rm S} K^0_{\rm S}$ Decays}
\end{center}
\vspace{0.9truecm}

\begin{center}
{\bf Robert Fleischer\,${}^{a,b}$,  Maria Laura Piscopo\,${}^{a,b}$, K. Keri Vos\,${}^{a,c}$ and \\
B. Ya\u{g}mur Zubaro\u{g}lu\,${}^{a}$}

\vspace{0.5truecm}

${}^a${\sl Nikhef, Science Park 105, NL-1098 XG Amsterdam,  Netherlands}

${}^b${\sl  Department of Physics and Astronomy, Vrije Universiteit Amsterdam,\\
NL-1081 HV Amsterdam, Netherlands}

{\sl $^c$Gravitational 
Waves and Fundamental Physics (GWFP),\\ 
Maastricht University, Duboisdomein 30,\\ 
NL-6229 GT Maastricht, Netherlands}\\[0.3cm]

\end{center}

\vspace*{1.7cm}


\begin{abstract}
\noindent
The non-leptonic $D^0\to K^- K^+$ and $D^0\to \pi^-\pi^+$ decays are powerful probes of the Standard Model and are related to each other through the $U$-spin symmetry of the strong interaction. 
Using lattice QCD inputs we calculate the corresponding colour-allowed tree amplitudes in factorisation and demonstrate that non-factorisable contributions and $U$-spin-breaking effects at the level of 50\% allow us to accommodate the measured branching ratios in the Standard Model. An exciting direct probe of such non-factorisable and $U$-spin breaking effects is provided by the $D^0\to K^0_{\rm S} K^0_{\rm S}$ channel. This decay is governed by non-factorisable exchange topologies and essentially vanishes in the $U$-spin limit, although it is experimentally well established with a prominent branching ratio. Extrapolating our $D^0\to K^- K^+$ results using the isospin symmetry, we find a consistent benchmark picture. Specifically, 
we can accommodate the measured  $D^0\to K^0_{\rm S} K^0_{\rm S}$ branching ratio with $U$-spin-breaking effects at the 50\% level and exchange amplitudes at the level of 50\% of the colour-allowed $D^0\to K^- K^+$, $D^0\to \pi^-\pi^+$ tree contributions. Finally, we explore the resulting range for direct CP violation in $D^0\to K^0_{\rm S} K^0_{\rm S}$, obtaining upper bounds in our benchmark scenarios of a few per mille, offering an exciting target for future measurements. 
\end{abstract}


\vspace*{2.1truecm}

\vfill

\noindent
December 2025

\end{titlepage}

\thispagestyle{empty}

\vbox{}

\setcounter{page}{0}

\newpage
\section{Introduction}\label{sec:intro}
The charm sector offers a valuable setting for studying and testing the quark-flavour sector of the Standard Model (SM) and the interplay with strong interaction dynamics (see e.g.~\cite{Friday:2025gpj} for a recent review). 
Of particular interest are the singly Cabibbo-suppressed~(SCS) decays
$D^0\to K^-K^+$ and $D^0\to \pi^-\pi^+$. Within the SM, they receive contributions from colour-allowed tree as well as exchange and penguin topologies. The latter enter with a strongly suppressed factor of Cabibbo--Kobayashi--Maskawa (CKM) matrix elements involving a CP-violating complex phase which can generate direct CP violation through interference with the colour-allowed tree and exchange amplitudes. 

Direct CP violation in the charm sector has indeed been established by the LHCb collaboration at the per-mille level through the measurement of the difference of CP asymmetries in the
$D^0\to K^-K^+$ and $D^0\to \pi^-\pi^+$ modes~\cite{LHCb:2019hro}. This result raises the question of whether such an effect is compatible with SM expectations, which naively predict a direct CP asymmetry about an order of magnitude smaller, or whether it might hint at contributions from physics beyond the SM~(BSM)~\cite{Khodjamirian:2017zdu, Grossman:2019xcj, Schacht:2021jaz, Chala:2019fdb, Lenz:2023rlq, Pich:2023kim, Bediaga:2022sxw, Dery:2019ysp, Cheng:2019ggx}. 
We will touch on this exciting question in this work.

For the branching ratios, the penguin contributions play a negligible role because of the tiny associated CKM factor. The measurements show the following pattern \cite{ParticleDataGroup:2024cfk}:
\begin{align} 
& {\cal B}(D^0 \to \pi^- \pi^+)|_{\rm exp} \,\,\,\,= (1.454 \pm 0.024) \times 10^{-3}\,,
\label{eq:Br_pipi_exp-0}\\[2mm]
& {\cal B}(D^0 \to K^- K^+)|_{\rm exp} = (4.08 \pm 0.06) \times 10^{-3}\,.
\label{eq:Br_KK_exp-0}
\end{align}
Interestingly, the decay topologies of $D^0\to K^-K^+$ and $D^0\to \pi^-\pi^+$ are related to one another through the $U$-spin symmetry of the strong interaction, in analogy to the $B^0_s\to K^-K^+$ and $B^0_d\to \pi^-\pi^+$ system~\cite{Fleischer:1999pa, Fleischer:2022rkm}. Because of the different CKM factors and the down-type quarks appearing in penguin topologies of charm decays, the phenomenology of $D$ decays differs significantly from $B$ decays, despite the similar $U$-spin relations in their decay topologies. In the exact $U$-spin limit the branching ratios of the $D^0\to K^-K^+$ and $D^0\to \pi^-\pi^+$ decays would be equal, whereas the values above differ substantially, indicating sizeable $U$-spin-breaking contributions.
To investigate if this pattern can be understood in the SM, in this work, we  
apply the factorisation framework to the colour-allowed tree amplitudes, expressing them in terms of the product of non-perturbative decay constants and form factors which are known from lattice QCD calculations. Interestingly, 
we find that our predictions for the factorised decay rates resemble the pattern of the measured branching ratios pretty well. 
Performing a detailed numerical analysis, we 
observe that the experimental data can be accommodated with non-factorisable and $U$-spin-breaking effects at the level of $50\%$. 
These corrections are larger than those in the $B^0_s\to K^-K^+$ and $B^0_d\to \pi^-\pi^+$ systems~\cite{Fleischer:2022rkm}. However, since the charm-quark mass $m_c\sim1.5$\,GeV is smaller than the $b$-quark mass $m_b\sim 5$\,GeV and non-factorisable effects are expected to scale as $\Lambda_{\rm QCD}/m_Q$ with the heavy-quark mass $m_Q$, this result is actually not surprising. 
The non-factorisable amplitude contains contributions from the exchange topologies, which do not factorise, but also non-factorisable effects related to the colour-allowed tree amplitude. Furthermore, we have a contribution from QCD penguin topologies, which is, however, suppressed by the difference of the strange- and down-quark masses (and would vanish in the $U$-spin limit).
Remarkably, we find SM benchmark scenarios in which the observed branching ratios can be accommodated without invoking anomalously large non-factorisable and $U$-spin-breaking effects.

In order to obtain further and complementary insight into the size of these contributions, we also analyse the decay $D^0\to K^0_{\rm S}K^0_{\rm S}$ (see also e.g., \cite{Nierste:2015zra, Pich:2023kim, Cheng:2019ggx}). This channel has a very intriguing dynamics as it receives only contributions from exchange and penguin-annihilation topologies, which are expected to be suppressed, and essentially vanishes in the strict $U$-spin limit. Interestingly, this channel is experimentally well established with the following branching ratio \cite{ParticleDataGroup:2024cfk}:
\begin{equation} 
{\cal B}(D^0 \to K_{\rm S}^0 K_{\rm S}^0)|_{\rm exp} = (1.41 \pm 0.05) \times 10^{-4}\,.
\label{eq:Br_KsKs_exp-0}
\end{equation}
The key question is therefore whether this value can be reproduced within the SM.

To address it, we make use of the information extracted from our analysis of the $D^0\to K^-K^+$ and $D^0\to \pi^-\pi^+$ decays. Assuming that the non-factorisable effects are governed by the exchange topologies, we can employ the isospin symmetry and use the $D^0\to K^-K^+$ decay to constrain one of the two exchange amplitudes of the $D^0 \to K_{\rm S}^0 K_{\rm S}^0$ channel. Then, taking into account the measured branching ratio in Eq.~(\ref{eq:Br_KsKs_exp-0}), we determine the other exchange amplitude and explore the size of the corresponding $U$-spin-breaking effects. We find again that effects at the level of 50\% can accommodate the data. Finally, we also calculate the direct CP asymmetry in the $D^0 \to K_{\rm S}^0 K_{\rm S}^0$ channel, finding upper bounds for our SM benchmark points up to the per-mille regime, providing interesting targets for future measurements. 
The most recent experimental determinations of the time-integrated CP asymmetry in this mode come from the LHCb~\cite{LHCb:2025ezf}, Belle II~\cite{Belle:2024vho} and CMS~\cite{CMS:2024hsv} collaborations. The resulting world average, including all available data~\cite{Punzi}, reads as: 
\begin{equation} 
    A_{\rm CP}(D^0\to K_{\rm S}^0 K_{\rm S}^0)|_{\rm exp} = (- 0.17 \pm 0.65)\%\,.
    \label{eq:ACPKSKS}
\end{equation}
This result, while statistically limited, provides an important benchmark for theoretical expectations and highlights the importance of improved measurements in the near future.

The outline of this paper is as follows: In Section~\ref{sec:pipi and KK}, we introduce the theoretical framework for the study of the $D^0\to K^-K^+$ and $D^0\to \pi^-\pi^+$ decays, while the detailed analysis of the patterns of the non-factorisable contributions following from the experimental data is discussed in Section~\ref{sec:nonfac}. Here we also show that the measured direct CP violation in the $D^0\to K^- K^+$, $D^0\to \pi^-\pi^+$ system requires significant strong penguin amplitudes in the ballpark of the tree amplitudes. In Section~\ref{sec:KSKS}, we present the analysis of the $D^0 \to K_{\rm S}^0 K_{\rm S}^0$ decay. Finally, we give our conclusions and a brief outlook in Section~\ref{sec:concl}.

\section{\boldmath Theoretical framework for studying the $D^0 \to \pi^-  \pi^+$ and $ D^0 \to K^-  K^+$ decays} 
\label{sec:pipi and KK}
\subsection{\boldmath Decay structure}
\begin{figure}
    \centering
\includegraphics[width=0.32\linewidth]{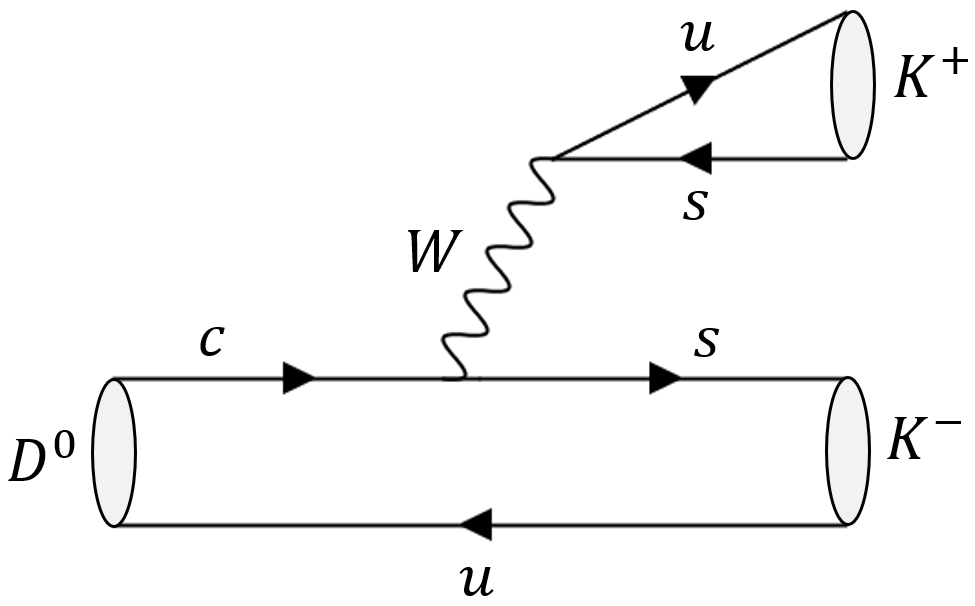}
    \,
\includegraphics[width=0.32\linewidth]{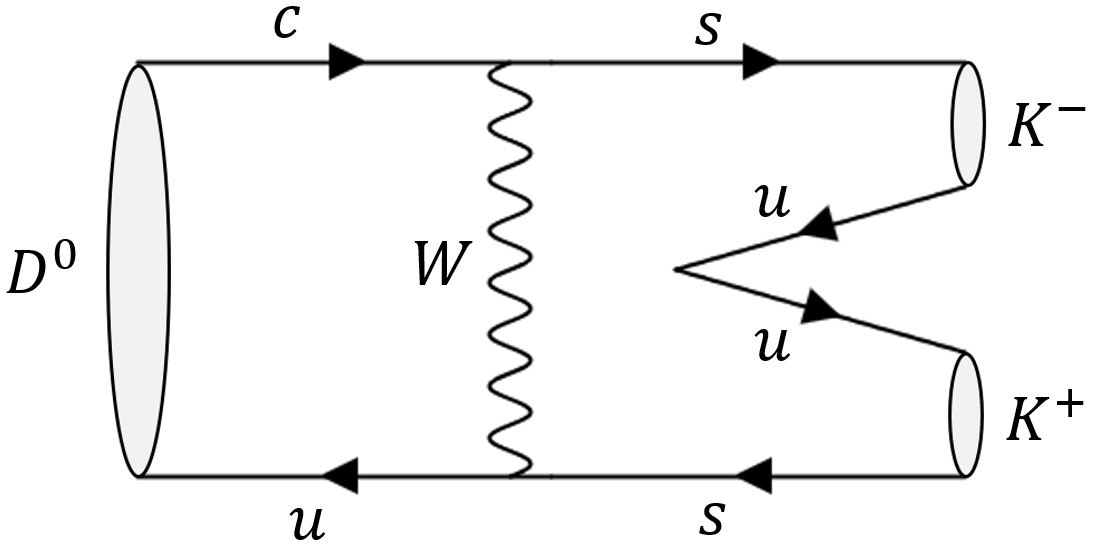}
    \,
\includegraphics[width=0.32\linewidth]{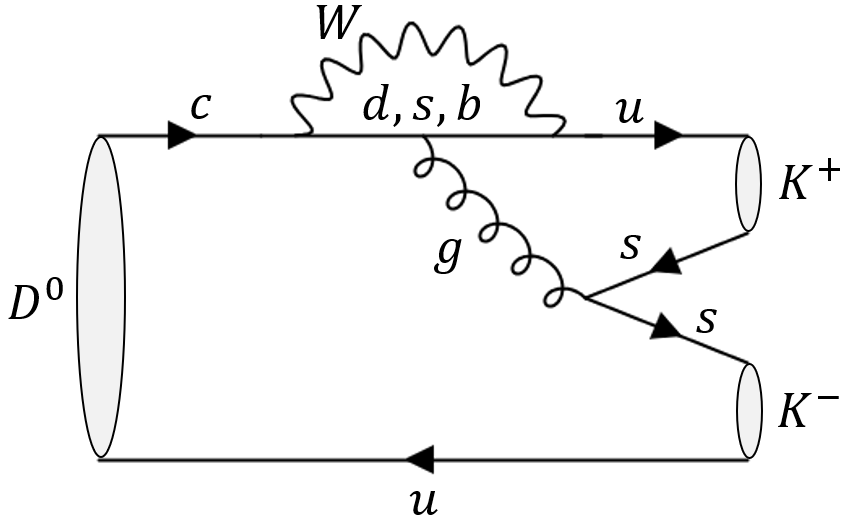}
    \caption{Examples of colour-allowed tree (left), exchange (middle), and penguin (right) topologies contributing in the SM to the decay amplitude of  $D^0 \to K^-K^+$. Corresponding diagrams for $D^0 \to \pi^- \pi^+$ are obtained by replacing$s \to d$ and $K \to \pi$.}
    \label{fig:DtoKK}
\end{figure}

We begin by considering SCS decays of the $D^0$ meson into final states with charged particles, namely $D^0\to \pi^- \pi^+$ and $D^0\to K^- K^+$. The first proceeds through the transition $c \to d \bar d u$, and the second through $c \to s \bar s u$.
Within the SM, these modes receive contributions from colour-allowed tree ($T$), exchange ($E$), and penguin ($P$) topologies~\footnote{The term penguin topology also includes penguin-annihilation contributions, which receive a further loop-suppression.}, as illustrated in Fig.~\ref{fig:DtoKK}.
The two decays are related to each other by the $U$-spin symmetry, which implies a one-to-one correspondence between their contributing topologies.

Specifically, we can express the decay amplitude of $D^0 \to K^- K^+$ as
\begin{equation} 
        A(D^0\to K^- K^+) = \lambda_s (T+ E + P_{s}) + \lambda_d P_d + \lambda_b P_{b} \,,
        \label{eq:kk1}
\end{equation}
where $\lambda_q \equiv V_{cq}^* V_{uq}$ with $q = d,s,b$ are factors with the corresponding elements of the CKM matrix, and $P_q$ denotes the penguin topology with an internal $q$ quark. 
Using the unitarity of the CKM matrix, which implies
\begin{equation}\label{eq:ckmuni} 
    \lambda_d + \lambda_s + \lambda_b = 0\,,
\end{equation}
we can write the decay amplitude in terms of only two independent CKM contributions:
\begin{equation}
A(D^0 \to K^-K^+) = \lambda_s \;{\cal A}_{KK} \left(1+ \frac{\lambda_b}{\lambda_s} a_{KK} \; e^{i\theta_{KK}}\right)\,.
\label{eq:AmplKK}
\end{equation}
Here 
    \begin{equation} 
     \mathcal{A}_{KK} \equiv T + E + P_{sd} \, , \qquad a_{KK} \; e^{i\theta_{KK}} \equiv \frac{P_{bd}}{ T + E + P_{sd}} \, ,
     \label{eq:AKK_fact}
\end{equation}
where $P_{qr} \equiv P_q - P_r$ denotes the difference of penguin amplitudes with internal quarks $q$ and $r$. 
A similar expression to the one in Eq.~\eqref{eq:kk1} holds for the $D^0 \to \pi^- \pi^+$ decay. In this case, we can use the CKM unitarity to eliminate $\lambda_s$, obtaining
\begin{equation} 
A(D^0 \to \pi^-\pi^+) = \lambda_d \;{\cal A}_{\pi\pi} \left(1+ \frac{\lambda_b}{\lambda_d} a_{\pi\pi}\; e^{i\theta_{\pi\pi}} \right)\,,
\label{eq:Amplpipi}
\end{equation}
where
\begin{equation} 
\mathcal{A}_{\pi\pi} \equiv T' + E' - P'_{sd} \, , \qquad a_{\pi\pi}\;e^{i\theta_{\pi\pi}} \equiv \frac{P'_{bs}}{T' + E' - P'_{sd}} \, ,
     \label{eq:Apipi_fact}
\end{equation}
with the primes indicating that we now have a $c \to d$ transition. 

Parametrisations similar to the ones above have been widely employed in the case of non-leptonic $B$-meson decays, see e.g.~\cite{Fleischer:1999pa, Fleischer:2022rkm}. However, in the charm system, the relevant CKM factors show a pronounced hierarchy, which becomes immediately transparent using the Wolfenstein parametrisation~\cite{Wolfenstein:1983yz}:
    \begin{align}\label{eq:wolf} 
    \lambda_s &= -\lambda_d + \mathcal{O}(\lambda^5) = \lambda \left(1 - \frac{\lambda^2}{2}\right) + {\cal O}(\lambda^5)\,, 
    \end{align}
where $\lambda \equiv |V_{us}|=0.225$ \cite{Charles:2004jd}. The two parameters $\lambda_s$ and $\lambda_s$ are thus opposite in sign and fully governed by $|V_{us}|$.~\footnote{Tiny differences between $\lambda_s$ and $\lambda_s$ arise only at ${\cal O}(\lambda^5)$ and also introduce an imaginary part for these coefficients.} 

The CKM factor $\lambda_b$ actually arises at ${\cal O}(\lambda^5)$: 
    \begin{align} 
    \lambda_b &=  \left(\frac{\lambda^5 }{1-\frac{\lambda^2}{2}}\right) A^2 R_b \;e^{-i\gamma} + {\cal O}(\lambda^{11})\,,
\end{align}
where $A \equiv |V_{cb}/\lambda^2| \sim 0.8$;

\begin{align}\label{eq:Rb} 
   & R_b \equiv \left( 1 - \frac{\lambda^2}{2} \right) \frac{1}{\lambda} \left| \frac{V_{ub}}{V_{cb}} \right|  \sim 0.4
\end{align}

is the side of the unitarity triangle (UT) from the origin to the apex in the complex plane; and $\gamma\sim 65^\circ$ is the angle between $R_b$ and the real axis~\cite{DeBruyn:2025rhk}. 
The parameter $\lambda_b$ has the largest relative imaginary part and therefore plays a  crucial role in CP violation studies in charm decays. We note that $\lambda_b$ depends on $|V_{cb}|$ and $|V_{ub}|$ which suffer from tensions between determinations from inclusive or exclusive semileptonic $B$ decays \cite{Gambino:2020jvv}. We will return to this point when discussing the CP asymmetry in $D^0 \to K^- K^+$ and $D^0 \to \pi^- \pi^+$ in Section~\ref{sec:cp}.

Numerically, we have $|\lambda_b/\lambda_{s,d}| \sim \lambda^4 \sim 10^{-3}$. Therefore, we refer to the quantities $\mathcal{A}_{KK}$ and $\mathcal{A}_{\pi\pi}$ in Eqs.~\eqref{eq:AmplKK} and \eqref{eq:Amplpipi} as the CKM-leading amplitudes. The parameters
$a_{KK}$ and $a_{\pi\pi}$ measure the relative magnitude of CKM-suppressed to CKM-leading contributions, where $\theta_{KK}$ and $ \theta_{\pi\pi}$ are the associated CP-conserving strong phases. 

The decay amplitudes of $D^0 \to K^- K^+$ and $D^0 \to \pi^- \pi^+$
exhibit a similar structure and differ only by the opposite sign of the light-quark penguin contributions $P_{sd}$ and $P_{sd}^{\prime}$, reflecting that these modes are $U$-spin partners. 
In the limit of vanishing light quark masses, the penguin amplitudes with internal $d$ and $s$ quarks coincide:
\begin{equation} 
    P_d = P_s \, , \quad \quad P_d^\prime = P_s' \,,
\end{equation}
implying $P'_{sd} = P_{sd} = 0 $. In addition, $U$-spin symmetry links the two decays, implying
\begin{equation} 
   \mathcal{A}_{KK} = \mathcal{A}_{\pi\pi} \,, 
    \qquad a_{KK} = a_{\pi\pi} \ , 
    \qquad 
    \theta_{KK} = \theta_{\pi\pi} \, .
\end{equation}

\subsection{Branching ratios and CP asymmetries}
For the decay of a $D^0$ meson into a final state $f$, the CP-averaged, time-integrated decay rate is defined as
\begin{equation} 
    \langle \Gamma(f)\rangle \equiv \frac12 \left[\Gamma(D^0 \to f) + \Gamma(\bar D^0 \to \bar f) \right]\,.
    \label{eq:CP_averaged_rate}
\end{equation}
We refer to the branching ratio obtained from Eq.~\eqref{eq:CP_averaged_rate} as the {\it theoretical} branching ratio.~\footnote{This quantity 
is independent of $D^0$--$\bar{D}^0$ mixing effects, which would in principle need to be taken into account if the rate were not time-integrated and CP-averaged (see \cite{DeBruyn:2012wk} for a discussion of analogous effects in $B_s^0$ decays).}
In the case of $D^0 \to K^-K^+$, using the parametrisation given in Eq.~\eqref{eq:AmplKK}, it follows that
\begin{equation} 
    \langle \Gamma(K^-K^+)\rangle \propto |\lambda_s|^2 |\mathcal{A}_{KK}|^2 \left[1+ 2 \left|\frac{\lambda_b}{\lambda_s}\right| a_{KK} \cos\theta_{KK} \cos\gamma + \left|\frac{\lambda_b}{\lambda_s}\right|^2 a^{2}_{KK}\right] \, ,
    \label{eq:CP-averaged-rate}
    \end{equation}
    with an analogous expression holding for the $D^0 \to \pi^-\pi^+$ mode with the replacement $\lambda_s \to \lambda_d$, $a_{KK} \to - a_{\pi\pi}$ and $\theta_{KK}\to \theta_{\pi\pi}$. 
    As the penguin contribution $a_{KK}$ in Eq.~\eqref{eq:CP-averaged-rate} is strongly CKM suppressed, it  
    can be safely neglected when computing the branching ratio.
The CP-averaged branching ratio thus reads
\begin{equation} 
{\cal B}(D^0 \to K^-K^+) = {\cal N}_{KK} |\lambda_s|^2 |\mathcal{A}_{KK}|^2 \; \tau(D^0) \,,
\label{eq:BR}
\end{equation}
where  $\tau(D^0)$ is the $D^0$-meson lifetime, $\lambda_s$ is given in the Wolfenstein parametrisation in Eq.~\eqref{eq:wolf} and ${\cal N}_{KK}$ denotes the corresponding two-body phase-space factor. For decays into two charged pseudoscalar particles $P_1^-$ and $P_2^+$, i.e.\ $f = P_1^- P_2^+$, this factor is given by
\begin{equation}\label{eq:curlyN} 
{\cal N}_{f} \equiv  \frac{ 1}{16 \pi m_{D^0}^3} \sqrt{\lambda (m_{D^0}^2, m_{P_1}^2, m_{P_2}^2)}\,,
\end{equation} 
with $\lambda(a,b,c)\equiv (a-b-c)^2 - 4 bc$ being the K\"allen function. 

In the case of $D^0 \to \pi^- \pi^+$, the corresponding expression for the branching ratio is obtained from Eq.~\eqref{eq:BR} by replacing $KK \to \pi\pi$ and $\lambda_s \to \lambda_d$. Consequently, the branching ratios  of the $D^0 \to \pi^- \pi^+$ and $D^0 \to K^- K^+$ decays are entirely governed by the Wolfenstein parameter $\lambda$. This is in contrast to $B$-meson decays, where tensions between the CKM parameters $|V_{cb}|$ and $|V_{ub}|$ already affect the branching ratios, see e.g.,~\cite{DeBruyn:2022zhw,Fleischer:2022klb}.

Contrary to the CP-averaged branching ratio, the $\lambda_b$ terms are in fact crucial for the CP asymmetry. 
 Introducing the direct CP asymmetry in $D^0 \to f$ as follows:
\begin{equation}\label{eq:acpdef} 
a_{\rm CP}^{\rm dir} (f) \equiv \frac{|{{ A}(D^0 \to f)}|^2 - |{ { A}(\bar D^0 \to \bar f)}|^2}{|{{ A}(D^0 \to f)}|^2 + |{ { A}(\bar D^0 \to \bar f)}|^2}\,,
\end{equation}
we obtain
\begin{equation}\label{eq:acpa} 
   a_{\rm CP}^{\rm dir}(K^-K^+)=
\frac{
2\,\left|\dfrac{\lambda_b}{\lambda_s}\right|\, a_{KK}\, \sin\theta_{KK}\,\sin\gamma
}{
1
+ 2\,\left|\dfrac{\lambda_b}{\lambda_s}\right|\, a_{KK}\, \cos\theta_{KK}\,\cos\gamma
+ \left|\dfrac{\lambda_b}{\lambda_s}\right|^{2}\, a_{KK}^2
}\, .
    \end{equation}
The analogous expression for
$D^0\to \pi^-\pi^+$ is obtained replacing $|\lambda_s| \to |\lambda_d|$ and $a_{KK} \to - a_{\pi\pi}$, reflecting the difference in sign in the CKM factor in \eqref{eq:wolf}, as well as $\theta_{KK}\to \theta_{\pi\pi}$. 
Neglecting subleading corrections of the order $|\lambda_b/\lambda_{s,d}|^2$,  
the two direct CP asymmetries satisfy the following relation in the $U$-spin limit:
\begin{equation}
       a_{\rm CP}^{\rm dir} (K^-K^+)  = -    a_{\rm CP}^{\rm dir} (\pi^-\pi^+) = \frac{\Delta    a^{\rm dir}_{\rm CP}}{2} \, ,
       \label{eq:Delta_acp}
\end{equation}
where we have defined the difference between the CP asymmetries as
\begin{equation}
    \Delta a^{\rm dir}_{\rm CP}\equiv a^{\rm dir}_{\rm CP}( K^- K^+) - a^{\rm dir}_{\rm CP}(\pi^-\pi^+) \, .
\end{equation}

\subsection{Factorisation of the leading amplitude}\label{sec:theo}
A complete calculation of the $D^0 \to K^-K^+$ and $D^0\to \pi^-\pi^+$ decay amplitudes is notoriously challenging due to the interference of several topologies and the importance of non-perturbative effects. However, a first estimate can be obtained by employing naive factorisation. While factorisation has been extensively and successfully used in the $B$-meson sector, particularly for colour-allowed tree amplitudes, its applicability to hadronic charm decays is a priori less robust. This is due to the comparatively smaller charm-quark mass, which enhances non-factorisable contributions. It is therefore especially instructive to explore the implications of factorisation in order to gain insight into the underlying structure of non-factorisable effects in charm decays.

We first focus on the contributions to the leading amplitudes $\mathcal{A}_{KK}$ and $\mathcal{A}_{\pi\pi}$. In the factorisation limit, only the colour-allowed tree amplitudes $T$ and $T^\prime$ contribute to ${\cal A}_{KK}$ and ${\cal A}_{\pi\pi}$ in Eqs.~\eqref{eq:AKK_fact} and~\eqref{eq:Apipi_fact}, respectively. We can therefore write
\begin{equation}\label{eq:deltanf} 
    T=T_{KK}^{\rm fac} + T^{\rm non-fac}_{KK}\,,  \qquad  T' = T_{\pi\pi}^{\rm fac} + T^{\rm non-fac}_{\pi\pi}\, ,
\end{equation}
where $T_{KK}^{\rm fac}$ and $T_{\pi\pi}^{\rm fac}$ denote the colour-allowed tree amplitudes in the {\it strict factorisation} limit while $T^{\rm non-fac}_{KK}$ and $T^{\rm non-fac}_{\pi\pi}$ absorb the remaining non-factorisable contributions. 
Using the low-energy weak effective Hamiltonian~\cite{Buchalla:1995vs}, see Appendix~\ref{app:1}, the factorisable part of the amplitude for the $D^0\to P_1^-P_2^+$ decay can be computed as 
\begin{equation} 
T^{\rm fac}_{f}\equiv i \frac{G_F}{\sqrt{2}}  a_1 f_{P_2} \left( m_{D^0}^2 - m_{P_1}^2 \right) f_0^{D \to P_1}(m_{P_2}^2) \,,
\label{eq:Tf}
\end{equation}
in terms of tree-level matrix elements of the current-current effective operators $Q_1$ and $Q_2$, defined in Eqs.~\eqref{eq:Q1} and \eqref{eq:Q2}. Here $f_{P_2}$ is the decay constant of the $P_2^+$ meson and $f_0^{D \to P_1}(q^2)$ the scalar form factor describing the $D^0 \to P_1^-$ hadronic amplitude as a function of the transferred momentum squared $q^2$. 
The factor $a_1$ in Eq.~\eqref{eq:Tf} is defined as
\begin{equation} 
a_1(\mu) \equiv C_2 (\mu) + \frac{C_1 (\mu)}{3} \,,
\end{equation}
where $C_1(\mu)$ and $C_2(\mu)$ are the Wilson coefficients of the current-current operators, which depend on the renormalisation scale $\mu \sim m_c$. Table~\ref{tab:a1} shows a comparison of the numerical values of $a_1(\mu)$ at leading order (LO) and next-to-leading order (NLO) in QCD for different choices of $\mu$, calculated from the Wilson coefficients given in Table~\ref{tab:WC} of Appendix~\ref{app:1}. 

The resulting dependence on $\mu$ is mild, which supports the factorisation of the colour-allowed tree amplitude as non-factorisable contributions must cancel this $\mu$ dependence.  
This stands in contrast to colour-suppressed tree amplitudes, whose substantially larger 
$\mu$ dependence implies correspondingly larger non-factorisable contributions. We note, however, that compared to the $B$-meson system, the $\mu$ dependence of the $a_1$ parameter is stronger in the charm system \cite{Buras:1994ij}.

\begin{table}[t]
\centering
\begin{tabular}{|c|c|c|c|}
\hline
$\mu$[GeV] &  1 &  1.5 &  3 \\ 
\hline\hline
$a_1(\mu)$ & 1.14 (1.10) & 1.08 (1.06) & 1.04 (1.03)\\
\hline
\end{tabular}
\caption{Results at LO (NLO) for the parameter $a_1 (\mu)$ for different values of the renormalisation scale $\mu$. These are based on the values of the Wilson coefficients given in Table~\ref{tab:WC}.}
\label{tab:a1}
\end{table}

Singling out the factorisable contribution, we express the leading CKM amplitude as follows:
\begin{equation} 
{\cal A}_{f} =  T_{f}^{\rm fac} + \tilde {\cal A}_f  = T_{f}^{\rm fac}  \left( 1 + r_{f} e^{i \delta_{f}} \right) 
\label{eq:Ampl-decomposition}
\end{equation}
for $f = KK$ or $f = \pi\pi$. Here, $\tilde {\cal A}_f$ collects all deviations from the naive factorisation limit, including non-factorisable corrections to the tree topology $T^{\rm non-fac}_{f}$ as well as the exchange $E^{(\prime)}$ and penguin $P_{sd}^{(\prime)}$ topologies. 
We will collectively refer to $\tilde {\cal A}_f$  as the {\it non-factorisable} part of the amplitude.

Within the effective Hamiltonian framework, the light-quark penguin contributions $P_s^{(\prime)}$ and $P_d^{(\prime)}$ originate from penguin contractions of the current–current operators and may therefore be affected by sizeable rescattering effects. Their difference, however, is suppressed by the Glashow--Iliopoulos--Maiani (GIM) mechanism, which enhances the relative importance of the exchange topology within $\tilde {\cal A}_f$.

Moreover, factoring out $T_f^{\rm fac}$, we have parametrised the non-factorisable contributions on the r.h.s.\ of Eq.~\eqref{eq:Ampl-decomposition} as
\begin{equation}\label{eq:rf} 
    r_f e^{i\delta_f}\equiv \frac{ \tilde {\cal A}_{f} }{ T_{f}^{\rm fac}} \,,
\end{equation}
where $r_f$ denotes the magnitude of the non-factorisable amplitude relative to the factorisable one, and $\delta_f$ the corresponding relative strong phase. Since $T_f^{\rm fac}$ does not carry a strong phase, $\delta_f$ directly represents the strong phase of $\tilde {\cal A}_f$. 

While $T_f^{\rm fac}$ receives sizeable {\it factorisable $SU(3)$-breaking corrections} from form factors and decay constants, as discussed in detail in Section~\ref{sec:nonfac},
also the non-factorisable amplitude $\tilde {\cal A}_f$ may break the $SU(3)$ symmetry. 
In the limit of exact $U$-spin symmetry, these contributions coincide,
\begin{equation} 
    \tilde{\cal{A}}_{KK} = \tilde{\cal{A}}_{\pi\pi} \,,
\end{equation}
and departures from this relation probe {\it non-factorisable $SU(3)$-breaking corrections}.

The size of such effects can be quantified studying the ratio of the non-factorisable amplitudes in $D^0 \to K^- K^+$ and $D^0 \to \pi^- \pi^+$:
\begin{equation}
\left| \frac{\tilde {\cal A}_{KK}}{\tilde {\cal A}_{\pi\pi}} \right| = \frac{r_{KK}}{r_{\pi\pi}} \left| \frac{T_{KK}^{\rm fac}}{T_{\pi\pi}^{\rm fac}} \right|\,,
\label{eq:U_spin_break}
\end{equation}
as a function of the $U$-spin breaking in their relative strong phases $\Delta$, i.e.
\begin{equation} 
    \Delta \equiv \delta_{KK} -\delta_{\pi\pi} \,.
    \label{eq:delta_shift}
\end{equation}

\section{\boldmath Non-factorisable contributions in $D^0 \to \pi^-\pi^+$ and $D^0 \to K^- K^+$}
\label{sec:nonfac}
\subsection{Results for the branching ratios in factorisation}
In the factorisation limit, the branching ratio defined in Eq.~\eqref{eq:BR} reduces to
\begin{equation} 
{\cal B}(D^0 \to K^-K^+)|_{\rm fac}  \equiv  \, {\cal N}_{KK}\, |\lambda_s|^2 \, |T_{KK}^{\rm fac}|^2\, \tau(D^0)\,,
\label{eq:br_facdef}
\end{equation}
with an analogous expression holding for $D^0\to \pi^-\pi^+$.  
To compute the factorisable amplitude $T_f^{\rm fac}$ we use values for the decay constants and hadronic form factors obtained by lattice QCD determinations, see the FLAG review~\cite{FLAG:2024oxs}.~\footnote{Determinations of semileptonic $D \to \{\pi,K\}$ form factors are also available from light-cone sum rule calculations~\cite{Khodjamirian:2009ys}. We do not use these results in our analysis, due to their comparatively large uncertainties.} 

The pion and kaon decay constants are known with sub-percent uncertainties:
\begin{equation} 
f_\pi = (0.1302 \pm 0.0008)  {\rm \, GeV}\,, \qquad
f_K =  (0.1557 \pm 0.0003) {\rm \, GeV} \,.
\end{equation}
For the $D \to \{\pi, K\}$ form factors, several lattice QCD studies are available in the literature; see~\cite{FLAG:2024oxs} for a comprehensive summary. However, currently, for both transitions, results using ensembles with $n_f = 2 + 1 + 1$ active quark flavours have only been published by the ETM~\cite{Lubicz:2017syv} and, more recently, by the FNAL/MILC~\cite{FermilabLattice:2022gku} collaborations.
Using the values quoted in~\cite{FermilabLattice:2022gku, Lubicz:2017syv}, we obtain, 
\begin{align}
&
f_0^{D \to \pi}(m_\pi^2) = (0.632 \pm 0.005)\,, 
\quad
f_0^{D \to K}(m_K^2) = (0.772 \pm 0.003)\,
\quad \mbox{(FNAL/MILC '22)} 
 \label{eq:FF_FNAL}
\\[1mm]
&
f_0^{D \to \pi}(m_\pi^2) = (0.614 \pm 0.035)\,,
\quad
f_0^{D \to K}(m_K^2) = (0.789 \pm 0.028)\,
\quad \mbox{(ETM '18)} \, .
\label{eq:FF_valsM}
\end{align}
The relative uncertainties are small in both cases: about $(4$--$6) \%$ for ETM and even about a factor of ten smaller in the case of FNAL/MILC, i.e.\ $(4$--$8)$\textperthousand. Interestingly, the latter also show smaller $SU(3)$-breaking effects compared to the older ETM determinations:
\begin{align}
    & \frac{f_0^{D\to K}(m_K^2)}{f_0^{D\to \pi}(m_\pi^2)} = 1.222 \pm 0.011 \, \quad \mbox{(FNAL/MILC '22)}\\[2mm]
    & \frac{f_0^{D\to K}(m_K^2)}{f_0^{D\to \pi}(m_\pi^2)} = 1.285 \pm 0.086\,\quad \mbox{(ETM '18)} \, .
\end{align}

Using the values of the decay constants and form factors given above, we obtain for the ratios of factorised amplitudes:
\begin{equation} 
     \frac{T_{KK}^{\rm fac}}{T_{\pi\pi}^{\rm fac}} = 1.37^{+ 0.10}_{- 0.08}\, \quad \mbox{(FNAL/MILC '22)} \ ,
    \qquad
     \frac{T_{KK}^{\rm fac}}{T_{\pi\pi}^{\rm fac}} = 1.44^{+0.14}_{-0.13}\, \quad \mbox{(ETM '18)} \ ,
     \label{T_ratios}
\end{equation}
where the central values have been obtained using LO results for the parameter $a_1$ at the scale $\mu = 1.5$ GeV, cf.\ Table~\ref{tab:a1}, and PDG inputs for the relevant meson masses as well as for the $D^0$-meson lifetime~\cite{ParticleDataGroup:2024cfk}. This amounts to factorisable $U$-spin breaking of around $40\%$ at the level of the decay amplitude. The uncertainties reflect variations of all inputs parameters within their quoted ranges, together with the renormalisation-scale variation $\mu \in [1,3]$ GeV, added in quadrature. Note that although the FNAL/MILC form factors are significantly more precise than the ETM ones, the uncertainties on $T_f^{\rm fac}$ are similar.
This follows from the $\mu$-dependence of the Wilson coefficients. In fact, the uncertainty due to the scale variation largely dominates the final error in the case of the FNAL/MILC input, whereas it is comparable in size with the uncertainty due to the variation of the form factors in the case of the ETM results. 

Finally, the branching ratios of the $D^0 \to K^-K^+$and $D^0 \to \pi^-\pi^+$ decays within factorisation are found as  
\begin{equation} 
\hspace*{1.7cm} 
\left \{ \begin{array}{c}
{\cal B}(D^0 \to K^- K^+)|_{\rm fac}    = (3.27^{+0.36}_{-0.27}) \times 10^{-3}\,
\\[2mm]
{\cal B}(D^0 \to \pi^-\pi^+)|_{\rm fac}   \,\,\, = (2.03^{+0.23}_{-0.17}) \times 10^{-3}\,
\end{array}
\right.
\qquad 
\mbox{(FNAL/MILC\;'22)}\
\label{eq:Br_fac_FNAL}
\end{equation}
and
\begin{equation}
\hspace*{-0.4cm}
\left\{
\begin{array}{c}
{\cal B}(D^0 \to K^- K^+)|_{\rm fac}    = (3.42^{+0.45}_{-0.37}) \times 10^{-3}\,
\\[2mm]
{\cal B}(D^0 \to \pi^-\pi^+)|_{\rm fac}   \,\,\, = (1.92^{+0.31}_{-0.27}) \times 10^{-3}\,
\end{array}
\right.
\qquad
\mbox{(ETM\;'18)}\ .
\label{eq:Br_fac_ETM}
\end{equation}
Here now the uncertainties include variations of all input parameters and of the renormalisation scale, as well as of the CKM matrix elements. For the latter, we use the standard parametrisation of the CKM matrix with input values from~\cite{Charles:2004jd}. As expected, the CKM uncertainties have only a minor numerical impact. 

Comparing the results in Eqs.~\eqref{eq:Br_fac_FNAL} and \eqref{eq:Br_fac_ETM} with the experimental measurements in Eqs.~\eqref{eq:Br_pipi_exp-0} and \eqref{eq:Br_KK_exp-0},
we observe that within the factorisation approximation the predicted branching ratios lie within the ballpark of the experimental values, as previously noted in~\cite{Lenz:2023rlq}. For both decays, the deviations of the factorisation predictions from the measurements are at the level of (20–30)$\%$, but with opposite signs. Specifically, using the central values in Eq.~\eqref{eq:Br_fac_FNAL}, reaching the experimental results would require, at the amplitude level, a negative shift of about $-16\%$ for $D^0 \to \pi^- \pi^+$ and a positive shift of about $+12\%$ for $D^0 \to K^-K^+$. In the case of Eq.~\eqref{eq:Br_fac_ETM}, the corresponding shifts are approximately $-13\%$ and $+10\%$, respectively. It is interesting to note that in the leading amplitude $\mathcal{A}_{KK}$ and $\mathcal{A}_{\pi\pi}$ in Eq.~\eqref{eq:AKK_fact} and Eq.~\eqref{eq:Apipi_fact}, respectively, the penguin topologies $P_{sd}^{(')}$ also enter with opposite sign, thereby showing the same pattern. 

Finally, the ratio of the branching ratios reads
\begin{align}
     &\frac{{\cal B}(D^0 \to K^-K^+)|_{\rm fac}}{{\cal B}(D^0 \to \pi^-\pi^+)|_{\rm fac}} = 1.61^{+ 0.24}_{- 0.18}\, \quad \mbox{(FNAL/MILC '22)}\label{B_ratios1}
    \\[2mm]
    & \frac{{\cal B}(D^0 \to K^-K^+)|_{\rm fac}}{{\cal B}(D^0 \to \pi^-\pi^+)|_{\rm fac}} = 1.79^{+0.36}_{-0.31}\, \quad \mbox{(ETM '18)}\ ,
     \label{B_ratios2}
\end{align}
where we assumed uncorrelated uncertainties on the form factors and decay constants. These ratios quantify $U$-spin breaking corrections due to form factors and decay constants entering the factorisable amplitudes in Eq.~\eqref{T_ratios} and space-space effects. Comparing with the experimental results, yields
\begin{equation} 
\frac{{\cal B}(D^0 \to K^-K^+)|_{\rm exp}}{ {\cal B}(D^0 \to \pi^- \pi^+)|_{\rm exp} }= 2.81\pm 0.06 \ ,
\end{equation}
where we have again assumed uncorrelated uncertainties. We observe that the experimental results show an even larger amount of $U$-spin breaking than our predictions in Eq.~\eqref{B_ratios1} and Eq.~\eqref{B_ratios2}. We will have a closer look at these effects in the following subsection. 

\subsection{Extracting non-factorisable contributions 
}
\begin{figure}[t]
\centering
\includegraphics[scale = 0.8]{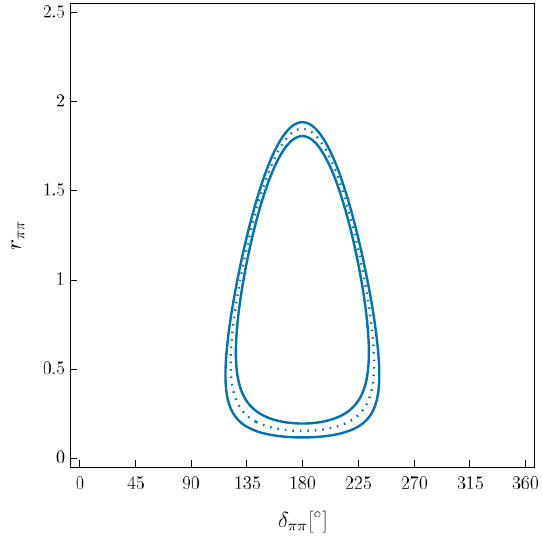}\qquad
\includegraphics[scale = 0.8]{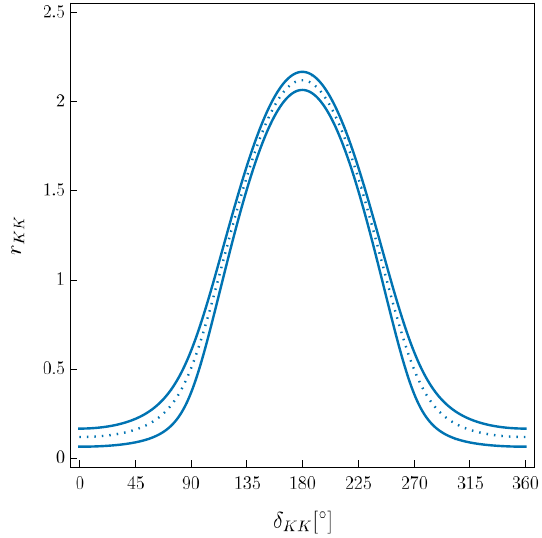}
\caption{Allowed ranges for the parameters $r_f$ and $\delta_f$ defined in Eq.~\eqref{eq:Ampl-decomposition} for $f = \pi\pi$~(left) and $f = KK$ (right). The curves have been obtained by fitting the expression in Eq.~\eqref{eq:Br-P1mP2p}, where the factorisation predictions are given in Eq.~\eqref{eq:Br_fac_FNAL}, to the corresponding experimental results in Eqs.~\eqref{eq:Br_pipi_exp-0} and \eqref{eq:Br_KK_exp-0}. In each plot, the dotted and solid lines indicate the contours obtained using the central and $1\sigma$ values for the factorisation results. Note that the effect of including also the experimental uncertainties falls within the shown curves.}
\label{fig:BrP1P2}
\end{figure}

Taking into account the parametrisation of the leading decay amplitude on the r.h.s.\ of Eq.~\eqref{eq:Ampl-decomposition}, and neglecting the tiny corrections proportional to $|\lambda_b/\lambda_{d,s}| \sim \lambda^4 \sim 10^{-3}$ in Eqs.~\eqref{eq:AmplKK} and \eqref{eq:Amplpipi}, the branching ratio, cf.\ Eq.~\eqref{eq:BR}, can be expressed compactly as follows
\begin{equation}
{\cal B}(D^0 \to f) = \Big[ 1 + r_{f}^2 + 2 r_{f} \cos \delta_{f} \Big] {\cal B}(D^0 \to f)|_{\rm fac} \,,
\label{eq:Br-P1mP2p}
\end{equation}
where $\cal{B}|_{\rm fac}$ is defined in Eq.~\eqref{eq:br_facdef} and we use the numerical  value in Eq.~\eqref{eq:Br_fac_FNAL}. We then  fit Eq.~\eqref{eq:Br-P1mP2p} to the corresponding experimental data to determine the allowed ranges for the magnitude $r_f$ and CP-conserving strong phase $\delta_f$ of the ratio of non-factorisable to factorisable amplitudes. The resulting contours in the $(\delta_f, r_f)$ plane are shown in Fig.~\ref{fig:BrP1P2} for $f = \pi \pi$ (left panel) and $f = K K$ (right panel).
As default, we use the most recent lattice QCD determinations of the form factors from FNAL/MILC, which also exhibit smaller uncertainties. However, the overall picture remains unchanged when using the ETM results. 

As seen in Fig.~\ref{fig:BrP1P2}, the allowed values of $r_f$ range from a few tenths up to about 2 for both modes. The qualitatively different shapes of the contours, i.e., closed for $D^0 \to \pi^- \pi^+$ and open for $D^0 \to K^-K^+$, reflect the fact that factorisation overshoots the data in the former case and undershoots it in the latter.
For $D^0 \to \pi^-\pi^+$, reproducing the measured branching ratio therefore requires destructive interference between the factorisable and non-factorisable contributions, implying
$\cos\delta_{\pi\pi}<0$. By contrast, for $D^0 \to K^- K^+$ the experimental branching ratio can be achieved either through constructive interference for small non-factorisable effects ($r_{KK} \lesssim 0.5$ with $\cos \delta_{KK} \geq 0 $), or through destructive interference if the non-factorisable amplitude is larger ($r_{KK} \gtrsim 0.5$ with $\cos \delta_{KK} \leq 0 $).

We emphasise that values as large as $r_f \sim 2$ cannot be mathematically excluded due to the structure of Eq.~\eqref{eq:Br-P1mP2p}. Indeed, the parameter combinations
\begin{equation}
r_f e^{i\delta_f} =
\left\{
\begin{array}{l}
\pm\, x_f\,  \\ 
- (2 \pm x_f)\,,
\end{array}
\right.
\,\,
\mbox{with}
\,\,\, 
x_f \ll 1 \,
\end{equation}
with the upper sign for $f=KK$ and the lower sign for $f=\pi\pi$, result in identical branching ratios and, thus, equally accommodate the data. Therefore, the appearance of large values of $r_f$ is simply a mathematical artifact rather than indicating genuinely large non-factorisable effects.

\begin{figure}
\centering
\includegraphics[scale=0.8]{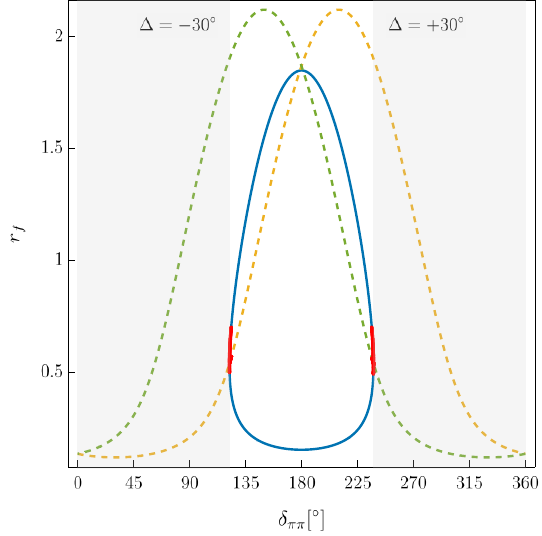}
\quad
\includegraphics[scale=0.8]{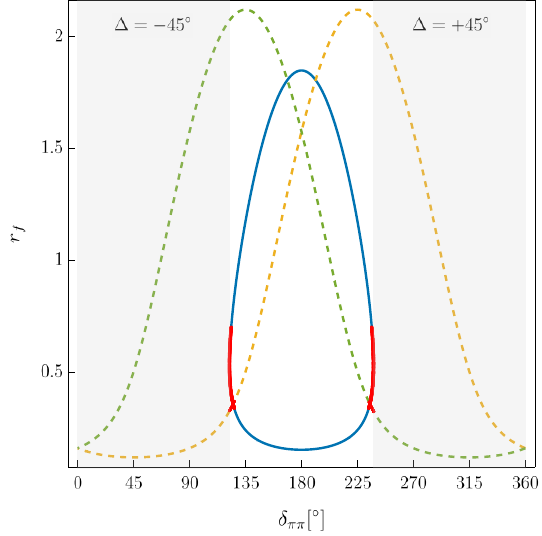}
\\[2mm]
\includegraphics[scale=0.8]{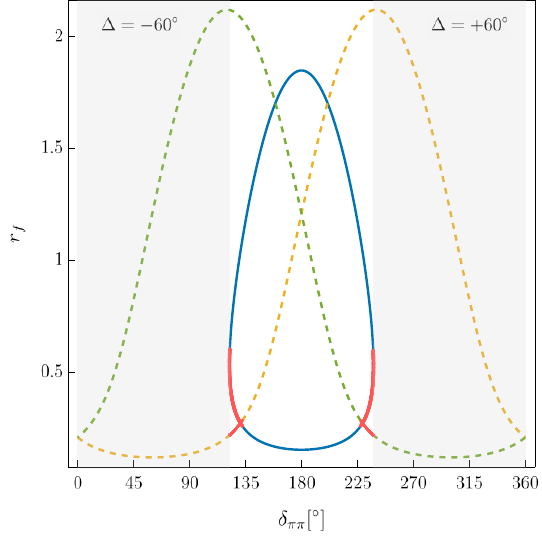}
\quad
\includegraphics[scale=0.8]{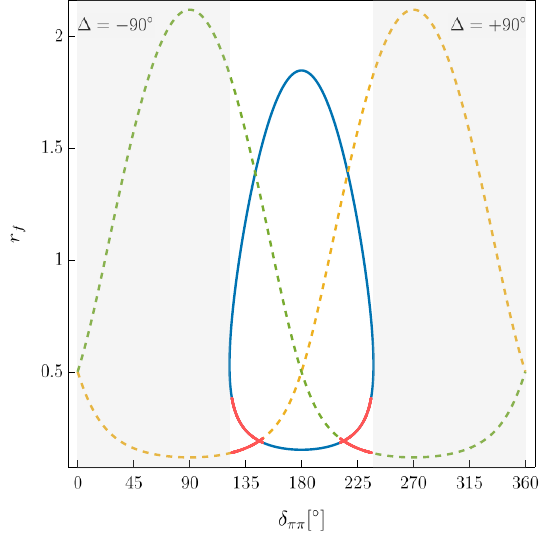}
\caption{Comparison of the contours for $r_f$ as a function of $\delta_{\pi\pi}$ for relative strong phase shifts $|\Delta|$ =  $30^\circ$ (top left), $45^\circ$ (top right), $60^\circ$ (bottom left) and $90^\circ$ (bottom right). In each plot, solid blue lines indicate the results for $r_{\pi\pi}$ while the dotted green and orange curves correspond to $r_{KK}$ for negative and positive values of
$\Delta$, respectively. The gray bands indicate the regions of $\delta_{\pi\pi}$ excluded by the analysis of Fig.~\ref{fig:BrP1P2}. The red highlighted regions indicate the parameter space satisfying the constraints in Eq.~\eqref{eq:constraints_P1P2}.
}
\label{fig:r_ratios}
\end{figure}

\begin{figure}
\centering
\includegraphics[scale=0.8]{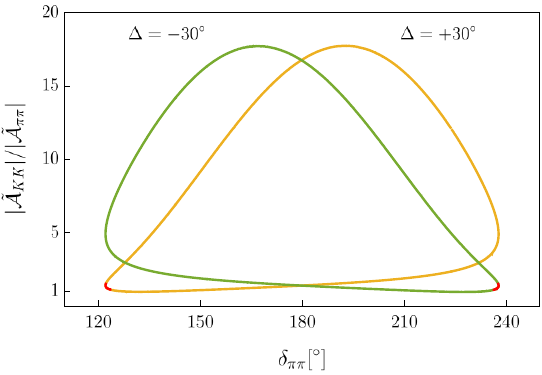}
\quad
\includegraphics[scale=0.8]{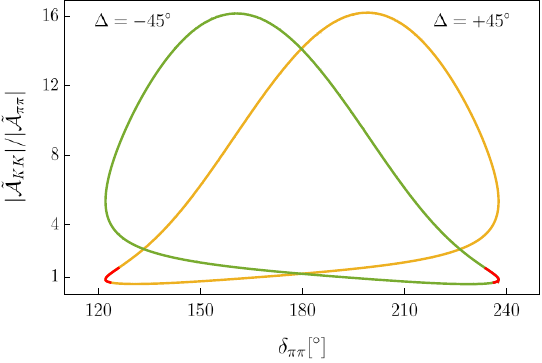}\\[2mm]
\includegraphics[scale=0.8]{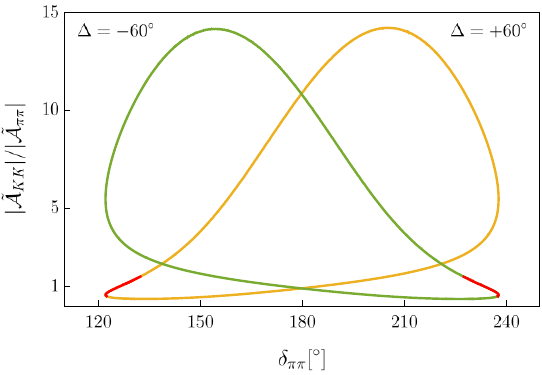}
\quad
\includegraphics[scale=0.8]{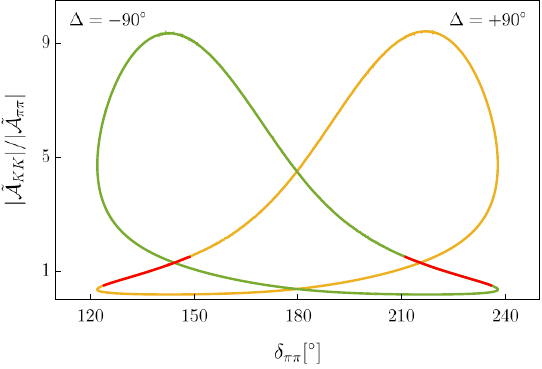}
\caption{Constraints on the size of the non-factorisable $U$-spin-breaking ratio $|\tilde {\cal A}_{KK}/\tilde {\cal A}_{\pi\pi}|$ as a function of the strong phase $\delta_{\pi\pi}$ for fixed relative strong-phase shifts $|\Delta|$ =  $30^\circ$ (top left), $45^\circ$ (top right), $60^\circ$ (bottom left) and $90^\circ$ (bottom right). In each plot the solid green and oranges lines correspond to negative and positive shifts, respectively. The red highlighted regions indicate the parameter space satisfying the constraints in Eq.~\eqref{eq:constraints_P1P2}.
}
\label{fig:A_ratios}
\end{figure}

\begin{figure}
\centering
\includegraphics[scale=0.8]{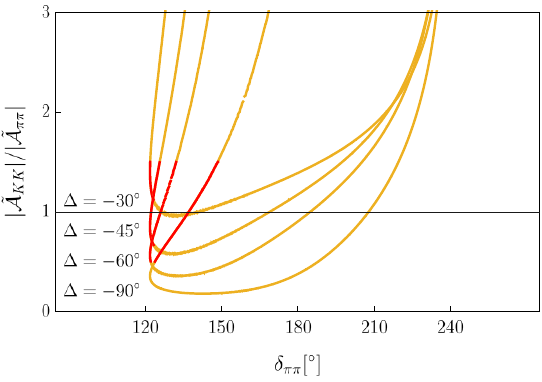}
\quad
\includegraphics[scale=0.8]{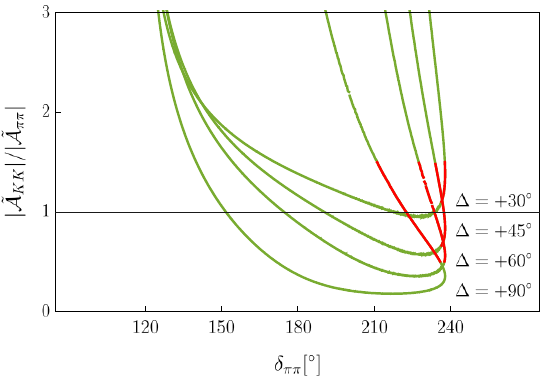}
\caption{Zoomed-in view of the allowed values of the non-factorisable $U$-spin-breaking ratio $|\tilde {\cal A}_{KK}/\tilde {\cal A}_{\pi\pi}|$ as a function of the strong phase $\delta_{\pi\pi}$ shown separately for negative (left) and positive (right) relative shifts $|\Delta|$ =  $30^\circ$, $45^\circ$, $60^\circ$, $90^\circ$. The horizontal gray line marks the $U$-spin symmetric limit. The red highlighted regions indicate the parameter space satisfying the constraints in Eq.~\eqref{eq:constraints_P1P2}.} 
\label{fig:AZoomed}
\end{figure}

Comparing the results for $D^0 \to \pi^- \pi^+$ and $D^0 \to K^-K^+$, we may obtain insights into the pattern of $U$-spin-breaking in the corresponding 
non-factorisable amplitudes. Interestingly, Fig.~\ref{fig:BrP1P2} suggests that very small values $r_f \ll 1$ tend to require large $U$-spin-breaking effects in the strong phases, pointing to a relative shift of about $180^\circ$.  
This can be quantified more precisely by studying the behaviour of $|\tilde {\cal A}_{KK}/\tilde{\cal A}_{\pi\pi}|$, given in Eq.~\eqref{eq:U_spin_break}, as a function of the strong-phase difference $\Delta$, defined in Eq.~\eqref{eq:delta_shift}. 

In Fig.~\ref{fig:r_ratios}, we show the contours for the central values $r_{\pi\pi}$ already given in Fig.~\ref{fig:BrP1P2} as a function of $\delta_{\pi\pi}$. In addition, we show the constraint on $r_{KK}$ as a function of $\delta_{\pi\pi}$ using fixed phase shifts $\Delta \equiv \delta_{KK}-\delta_{\pi\pi} = \pm (30^\circ, 45^\circ, 60^\circ, 90^\circ)$. The gray bands indicate the regions of $\delta_{\pi\pi}$ excluded by the analysis in Fig.~\ref{fig:BrP1P2}. 
For each chosen $\Delta$, moving along these contours, we can extract the corresponding value of $|\tilde {\cal A}_{KK}/\tilde{\cal A}_{\pi\pi}|$ as a function of $\delta_{\pi\pi}$, leading to the curves shown in Fig.~\ref{fig:A_ratios}.
We find that the implied $U$-spin-breaking effects in the non-factorisable amplitudes can, in principle, be extremely large, reaching even values of ${\cal O}(10)$ which are clearly unphysical. However, moderate values of order unity are also always allowed. A zoomed-in view of the physically reasonable region is shown in Fig.~\ref{fig:AZoomed}.

Can we actually find simultaneous ranges for all parameters with both moderate non-factorisable and moderate $U$-spin-breaking effects, allowing us to accommodate the measured branching ratios in the SM? In order to explore this key question, we consider the following benchmark ranges:
\begin{equation}
    r_{KK} \leq 0.7\,, \quad r_{\pi\pi} \leq 0.7\,, \quad
   0.5 \leq  \left| \frac{\tilde {\cal A}_{KK}}{\tilde {\cal A}_{\pi\pi}} \right| \leq 1.5\,.
    \label{eq:constraints_P1P2}
\end{equation}
The regions satisfying these conditions are highlighted in red in Figs.~\ref{fig:r_ratios}, ~\ref{fig:A_ratios} and~\ref{fig:AZoomed}. Although the allowed parameter space is significantly reduced, a solution is possible for all chosen values of $\Delta$. Interestingly, the values of $r_{KK}$ are particularly constrained and -- for larger values of $|\Delta|$ -- limited to only a few tenths.

\begin{figure}
    \centering
    \includegraphics[width=0.7\linewidth]{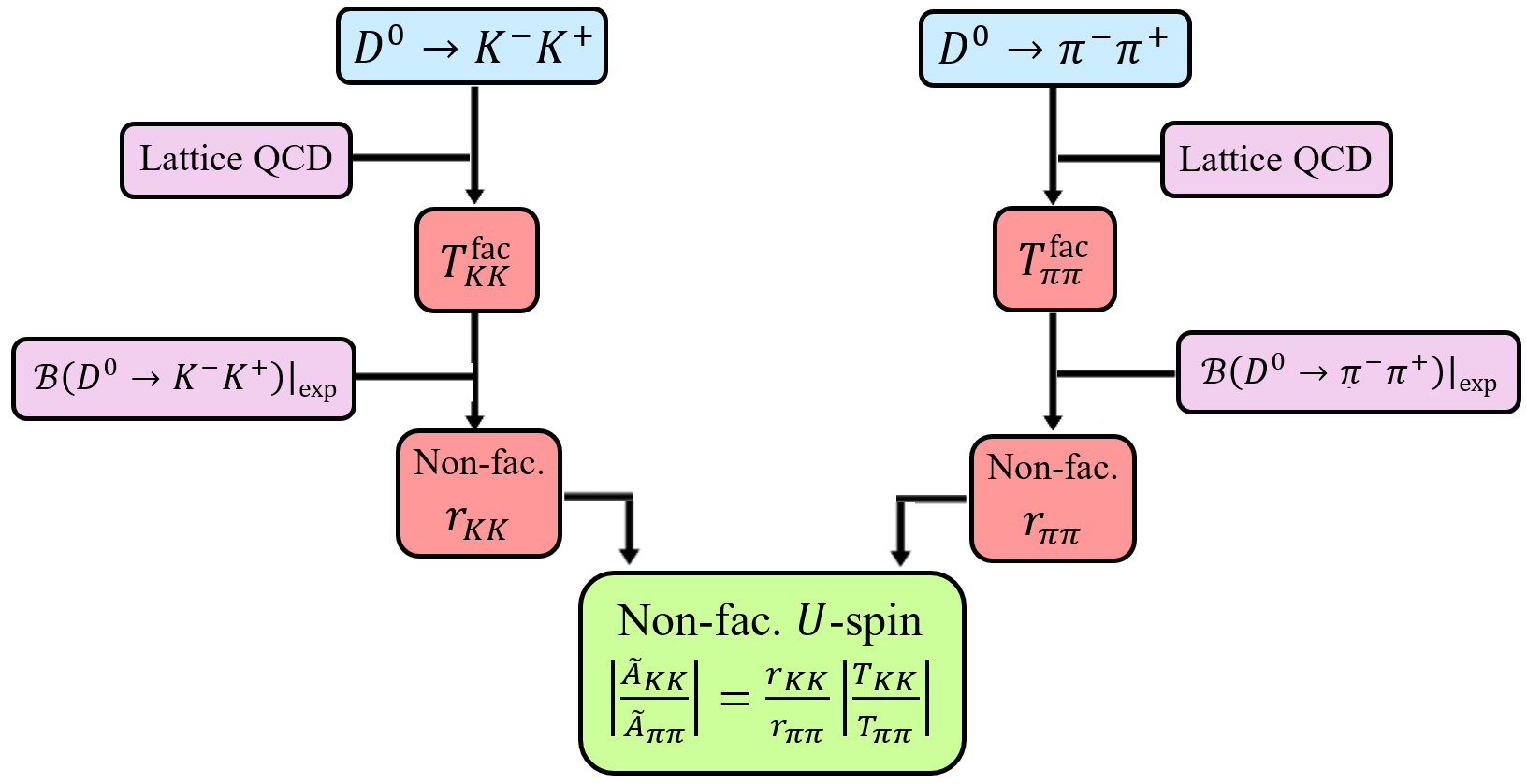}
    \caption{Overview of our strategy to study the SM benchmarks for the $D^0\to K^- K^+$ and $D^0\to \pi^-\pi^+$ decays.}
    \label{fig:charged flowchart}
\end{figure}

We summarize our analysis method in Fig.~\ref{fig:charged flowchart}. 
From our new analysis, we conclude that reasonable non-factorisable contributions with moderate $U$-spin breaking effects within these non-factorisable parameters, both up to $50\%$, can accommodate the pattern of the branching ratios in Eq.~\eqref{eq:Br_KK_exp-0} and \eqref{eq:Br_pipi_exp-0}. The size of this effect -- generally a bit larger than found in the $B$ system -- falls in line with expectations, considering also the lighter charm mass. We stress that these $U$-spin breaking effects enter only through already subleading non-factorisable topologies.
As such, their overall effect is cumulative.  

%
%
%
%
%
\subsection{Constraints  from the direct CP asymmetries}
\label{sec:cp}
In this section, we discuss the implications that the results in Fig.~\ref{fig:BrP1P2} have for the direct CP asymmetries in $D^0 \to \pi^- \pi^+$ and $D^0 \to K^- K^+$. 
Recall that in the $U$-spin limit, where $a_{KK}= a_{\pi\pi}$, and neglecting terms of order $|\lambda_b/\lambda_{d,s}|^2$, the difference of CP asymmetries given in Eq.~\eqref{eq:Delta_acp} can be approximated as
\begin{equation}
    \Delta a^{\rm dir}_{\rm CP} \sim 4 a_{KK} \left|\frac{\lambda_b}{\lambda_s}\right| \sin\theta \sin\gamma \sim 3 \times 10^{-3}  \left| \frac {P_{b}-P_q}{T+E} \right|\sin\theta \, ,
\end{equation} 
where we have used $\sin\gamma \sim 0.9$, and $P_q = P_d= P_s$. 
Comparing this with the measurement~\cite{LHCb:2019hro}
\begin{equation} 
    \Delta a^{\rm dir}_{\rm CP}|_{\rm exp} = (-15.7 \pm 2.9) \times 10^{-4} \,
    \label{eq:Delta_acp_exp}
\end{equation}
yields 
\begin{equation} 
    \left| \frac{P_{b} - P_{q}}{T+E} \right|  \sin \theta \sim -0.5 \, .
    \label{eq:Delta_acp_U_spin}
\end{equation}
This finding is highly non-trivial: it suggests no suppression of the penguin amplitude with respect to the colour-allowed tree and exchange topologies, thereby corresponding to an enhancement of roughly an order of magnitude compared to naive perturbative expectations. Whether this can be accommodated within the SM remains actively debated~\cite{Khodjamirian:2017zdu, Grossman:2019xcj, Schacht:2021jaz, Chala:2019fdb, Lenz:2023rlq, Pich:2023kim, Bediaga:2022sxw, Dery:2019ysp, Cheng:2019ggx}.

The most recent determination of the CP asymmetry in $D^0 \to K^- K^+$~\cite{LHCb:2022lry} has also raised new questions. Combining this result with $\Delta a_{\rm CP}^{\rm dir}$, the LHCb collaboration obtained~\cite{LHCb:2022lry}:~\footnote{For the most recent determination of these CP asymmetries, including all available data, see  see~\cite{LHCb:2024yxi}.} 
\begin{align} 
a_{\rm CP}^{\rm dir}(\pi^-\pi^+)|_{\rm exp} = (23.2 \pm 6.1)\times 10^{-4}\,,
\label{eq:acp_pipi_exp}
\\[1.5mm]
a_{\rm CP}^{\rm dir}(K^- K^+)|_{\rm exp} = (\phantom{1}7.7 \pm 5.7)\times 10^{-4}\,,
\label{eq:acp_KK_exp}
\end{align}
indicating a departure from the $U$-spin symmetry limit in Eq.~\eqref{eq:Delta_acp} of $2.7\sigma$. Although the current experimental precision does not yet permit firm conclusions, possible interpretations of the results in Eqs.~\eqref{eq:acp_pipi_exp} and \eqref{eq:acp_KK_exp} have been explored e.g.\ in~\cite{Schacht:2022kuj, Bause:2022jes, Iguro:2024uuw}. 
Furthermore, Eq.~\eqref{eq:acp_pipi_exp} provides the first evidence of CP violation in a specific $D^0$ decay, and its implications have been investigated e.g.\ in~\cite{Gavrilova:2023fzy, Sinha:2025cuo}.

To shed further light on this puzzling pattern, we now aim to better quantify the size of the penguin contributions $a_{KK}$ and $a_{\pi\pi}$, using again factorisation, as a starting point.
First, 
we write the amplitude of $D^0 \to K^- K^+$ as follows
\begin{equation}
A(D^0 \to K^-K^+) = \lambda_s  \left[T_{KK}^{\rm fac} + \tilde{{\cal A}}_{KK} \right]+ \lambda_b \left[P_{KK}^{\rm fac}+ \tilde{{\cal P}}_{KK} \right] \, ;
\label{eq:Amplfac}
\end{equation}
an analogous expression for
$D^0\to \pi^-\pi^+$ can be obtained by simply replacing
$\lambda_s\to \lambda_d$ and $KK \to \pi\pi$. The leading CKM amplitude was discussed in the previous section. For the subleading amplitude, $P_{KK}^{\rm fac}$ denotes the contribution of the QCD penguin operators computed in the factorisation limit. For the $D^0\to P_1^-P_2^+$ decay, at LO, this reads~\footnote{The minus sign originates from the overall sign of the QCD penguin operators
$Q_k$, with $k = 3, \ldots, 6$ in the effective Hamiltonian in Eq.~\eqref{eq:Eff_Ham}.}
\begin{equation}
P^{\rm fac}_f \equiv
-i \frac{G_F}{\sqrt{2}} \left[ \left( C_4 + \frac{C_3}{3} \right) + \frac{2 \mu_{P_2}}{m_c}  \left( C_6 + \frac{C_5}{3} \right) \right] f_{P_2} \left( m_{D^0}^2 - m_{P_1}^2 \right) f_0^{D \to P_1}(m_{P_2}^2)\,,
\label{eq:Pf}
\end{equation}
where $\mu_{P_2}$ is a chirally enhanced parameter with $\mu_{K} \equiv m_{K}^2/(m_u + m_s)$ and $\mu_{\pi} \equiv m_{\pi}^2/(m_u + m_d)$~\footnote{These mass factors follow from using the equations of motion for the quark fields in the parametrisation of the penguin operators $Q_{5}$ and $Q_6$ defined in Appendix~\ref{app:1}.}. 
In Eq.~\eqref{eq:Amplfac}, $\tilde{\mathcal{P}}_{KK}$ absorbs all remaining non-factorisable contributions, both from the QCD penguin operators and from penguin contractions of the current-current operators. For consistency, the Wilson coefficients in Eq.~\eqref{eq:Pf} must therefore be taken at LO~\cite{Fleischer:1992gp}. 
We note that the contribution of the QCD penguin operators is often neglected in literature due to their small Wilson coefficients, see, e.g.,~\cite{Lenz:2023rlq}. Under this assumption, $P_{KK}^{\rm fac} = 0$ and $\tilde {\cal P}_{KK}$ arises solely from penguin contractions of the current--current operators.

\begin{table}[t]
\centering
\begin{tabular}{|c|c|c|c|}
\hline
$\mu$\;[GeV] &  1 &  1.5 &  3 \\ 
\hline\hline
$c_{\pi\pi}(\mu)$ & 0.18 (0.30) & 0.10 (0.17) & 0.02 (0.06)\\
\hline
\end{tabular}
\caption{Comparison of the numerical values of the parameter $c_{KK}$, cf.\ Eq.~\eqref{eq:cf}, at LO (NLO) for different choices of the renormalisation scale $\mu$. The results are shown for the central values of the chirally enhanced parameter $\mu_{\pi} = 2.49$ and of the charm-quark mass $m_c = 1.27$ GeV in the $\overline{\rm MS}$ scheme. At the current level of accuracy, the same results also apply to $c_{\pi\pi}$, with $\mu_{K} = 2.50$.}
\label{tab:cf}
\end{table}

Isolating the contribution of $T_{f}^{\rm fac}$, we parametrise the full decay amplitude as 
\begin{equation}
A (D^0 \to K^-K^+) = \lambda_s T_{f}^{\rm fac}  \left[ \left( 1 + r_{KK} e^{i \delta_{KK}} \right) + \frac{\lambda_b}{\lambda_s} \left(c_{KK} + r^p_{KK} e^{i \delta^p_{KK}} \right) \right] \,,
\label{eq:Ampl-decomposition-KK}
\end{equation}
and analogously for $D^0\to \pi^-\pi^+$.  Here, $r_f$ and $\delta_f$
are defined in Eq.~\eqref{eq:rf}. For the CKM subleading amplitude, we now define
\begin{equation}
r^p_{f} e^{i\delta_f^p}\equiv \frac{ \tilde {\cal P}_{f} }{ T_{f}^{\rm fac}} \,.
\end{equation}
We stress that this quantity should not be confused with the parametrisation introduced in Eqs.~\eqref{eq:AmplKK} and \eqref{eq:Amplpipi}, where $a_{f} e^{i \theta_f}$ denotes the ratio of the full penguin contribution to the full leading amplitude.
Finally, in Eq.~\eqref{eq:Ampl-decomposition-KK}, we have defined
\begin{equation}
    c_f \equiv   - \left[ \left( C_4 + \frac{C_3}{3} \right) + \frac{2 \mu_{P_2}}{m_c}  \left( C_6 + \frac{C_5}{3} \right) \right] \left( C_2 + \frac{C_1}{3} \right)^{-1} \,.
    \label{eq:cf}
\end{equation}
A comparison of the values of this parameter for different choices of $\mu$ is given in Table~\ref{tab:cf}, in correspondence of the Wilson coefficients in Table~\ref{tab:WC} as well as $\mu_\pi= 2.5$ and $\mu_K = 2.49$~\cite{Khodjamirian:2017fxg}, and $m_c = 1.27$ GeV in the $\overline{\rm MS}$ scheme.
In our numerical analysis, we use $\mu=1.5$ GeV as a default, and 
\begin{equation}
    c_{\pi\pi} = c_{KK} = 0.1 \, , 
\end{equation}
while accounting for the uncertainties by varying the scale $\mu$ between $(1$--$3)$ GeV.

Taking into account the parametrisation of the decay amplitude given in Eq.~\eqref{eq:Ampl-decomposition-KK}, the CP asymmetry defined in \eqref{eq:acpa} can then be written as 
\begin{equation}
    a_{\rm CP}^{\rm dir}(f) = \zeta_f \, 2\left| \frac{\lambda_b}{\lambda_f} \right| \sin \gamma \left| \frac{c_f + r_f^p e^{i \delta_f^p} }{1 + r_f e^{i \delta_f}} \right| \sin \left[ \arg \left( \frac{c_f + r_f^p e^{i \delta_f^p} }{1 + r_f e^{i \delta_f}} \right) \right] \, \ ,
    \label{eq:acp_PpPm}
\end{equation}
where $\lambda_f = \lambda_s (\lambda_d)$ and $\zeta_f = 1 (-1)$ for $f = KK (\pi\pi)$. For simplicity, we give in Eq.~\eqref{eq:acp_PpPm} the approximate expression of the CP asymmetry obtained neglecting terms of order ${\cal O} (|\lambda_b/\lambda_{d,s}|^2)$. However, the full expression is used in our numerical analysis.

As the values of $r_f$ and $\delta_f$ are not fixed, we cannot constrain the parameter space for $r_f^p$ and $\delta_f^p$. However, for each point in the $\delta_f$--$r_f$ plane in Fig.~\ref{fig:BrP1P2}, the CP asymmetry in Eq.~\eqref{eq:acp_PpPm} becomes only a function of the two unknown parameters $r_f^p$ and $\delta_f^p$. Moving along the contours in Fig.~\ref{fig:BrP1P2}, and fitting the corresponding expression in Eq.~\eqref{eq:acp_PpPm} to the experimental determinations of the CP symmetries in Eqs.~\eqref{eq:acp_pipi_exp} and \eqref{eq:acp_KK_exp}, we can extract the minimum value of $r_f^p$ required to accommodate the experimental results, finding
\begin{equation} 
r_{\pi\pi}^p\big|_{\rm min} = (1.6 \pm 0.5)\,, \qquad 
r_{KK}^p\big|_{\rm min}  = (0.7 \pm 0.6)\,,
\label{eq:rp}
\end{equation}
where the uncertainties have been obtained by varying the input parameters such as the CKM matrix elements, the factor $c_f$, etc., as well as the experimental values of the CP asymmetries, within their error ranges. We find that at the moment the final uncertainties are dominated by the experimental ones. The strong phase $\delta_f$ remains largely unconstrained, as expected, since we only have one observable at our disposal.

From Eq.~\eqref{eq:rp}, sizeable values $r_{\pi\pi}^p \gtrsim 1$ are required to reproduce the CP asymmetry in the $D^0 \to \pi^- \pi^+$ channel, indicating a significant enhancement of the penguin contributions compared to the level of the colour-allowed tree amplitudes. More moderate values $r_{\pi\pi}^p \gtrsim 0.1$, consistent with naive estimates, are allowed at $3\sigma$. For the $D^0 \to K^- K^+$ mode, values $r_{KK}^p \gtrsim 0.1$ are consistent with the experimental data at $1\sigma$.

Finally, we note that the CP asymmetries are proportional to $\lambda_b$, which depends on the CKM matrix elements $|V_{cb}|$ and $|V_{ub}|$. Here, the situation is more involved than for the branching ratios, due to larger uncertainties and deviations in different determinations of these CKM parameters.  
However, current uncertainties, including differences between inclusive and exclusive determinations, see e.g.~\cite{Gambino:2020jvv, Bernlochner:2024sfg}, do not affect the analysis at the present level of precision, as our results are dominated by the experimental uncertainties on the CP asymmetries. Yet, we stress that the difference between the inclusive and exclusive determinations of these CKM matrix elements does not fall into the $1\sigma$ uncertainty range of $\lambda_b$. As such, taking the different CKM inputs may shift the values of the penguin parameters. Given the current uncertainties, this is not relevant at the moment, but it should be considered when the experimental determinations of the CP asymmetries become more precise. Of course, at that time, the inclusive/exclusive puzzle will also hopefully have been solved.


\section{\boldmath The decay $D^0 \to K_{\rm S}^0 K_{\rm S}^0$} \label{sec:KSKS}
\subsection{Decay structure and branching ratio}
\begin{figure}
    \centering
    \includegraphics[width=0.32\linewidth]{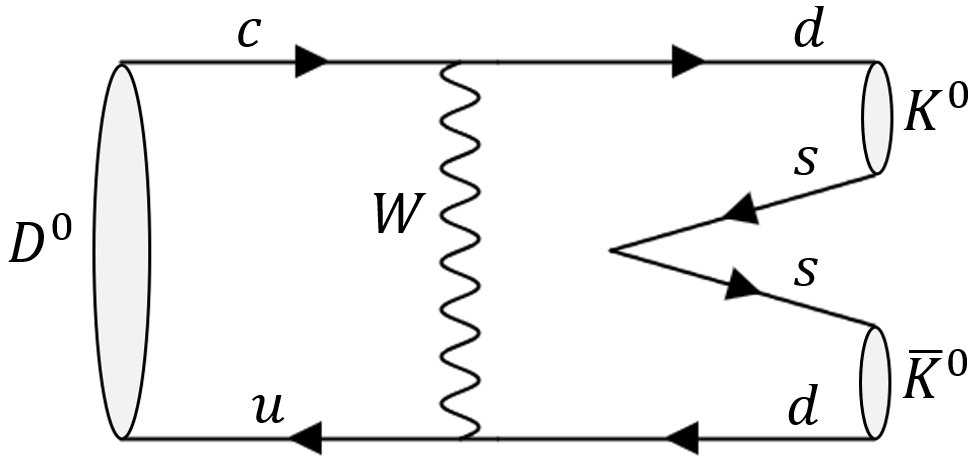} \,
    \includegraphics[width=0.32\linewidth]{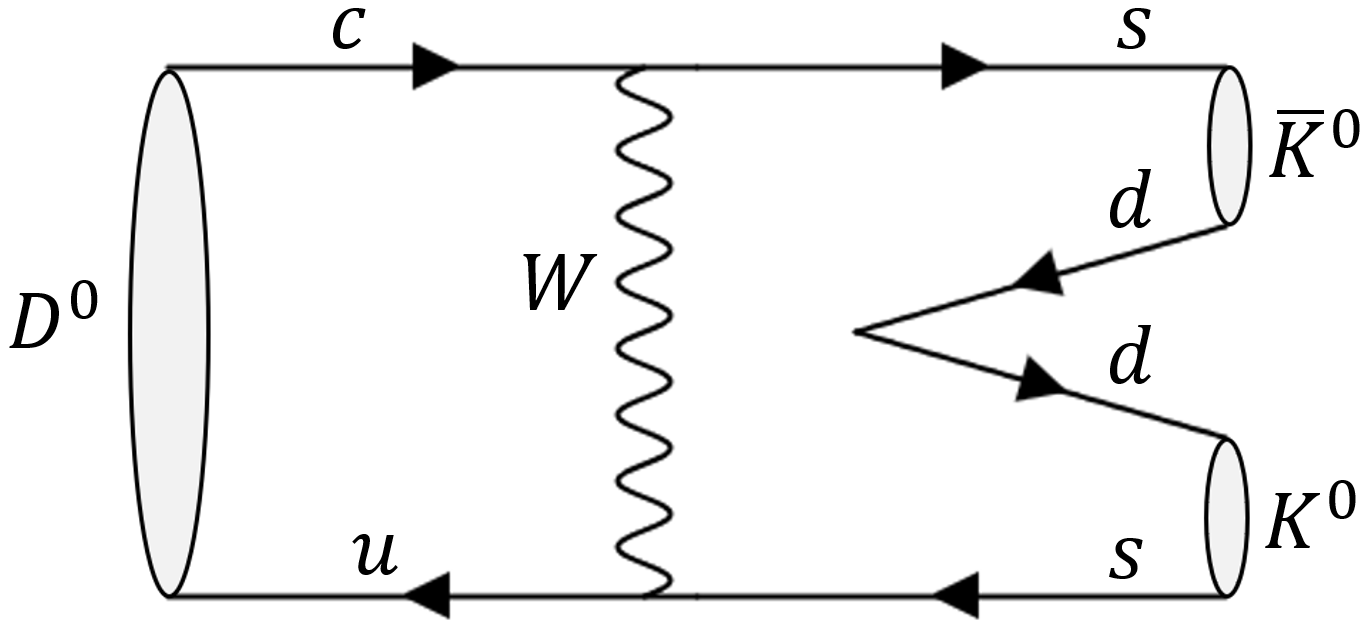} \,
    \includegraphics[width=0.32\linewidth]{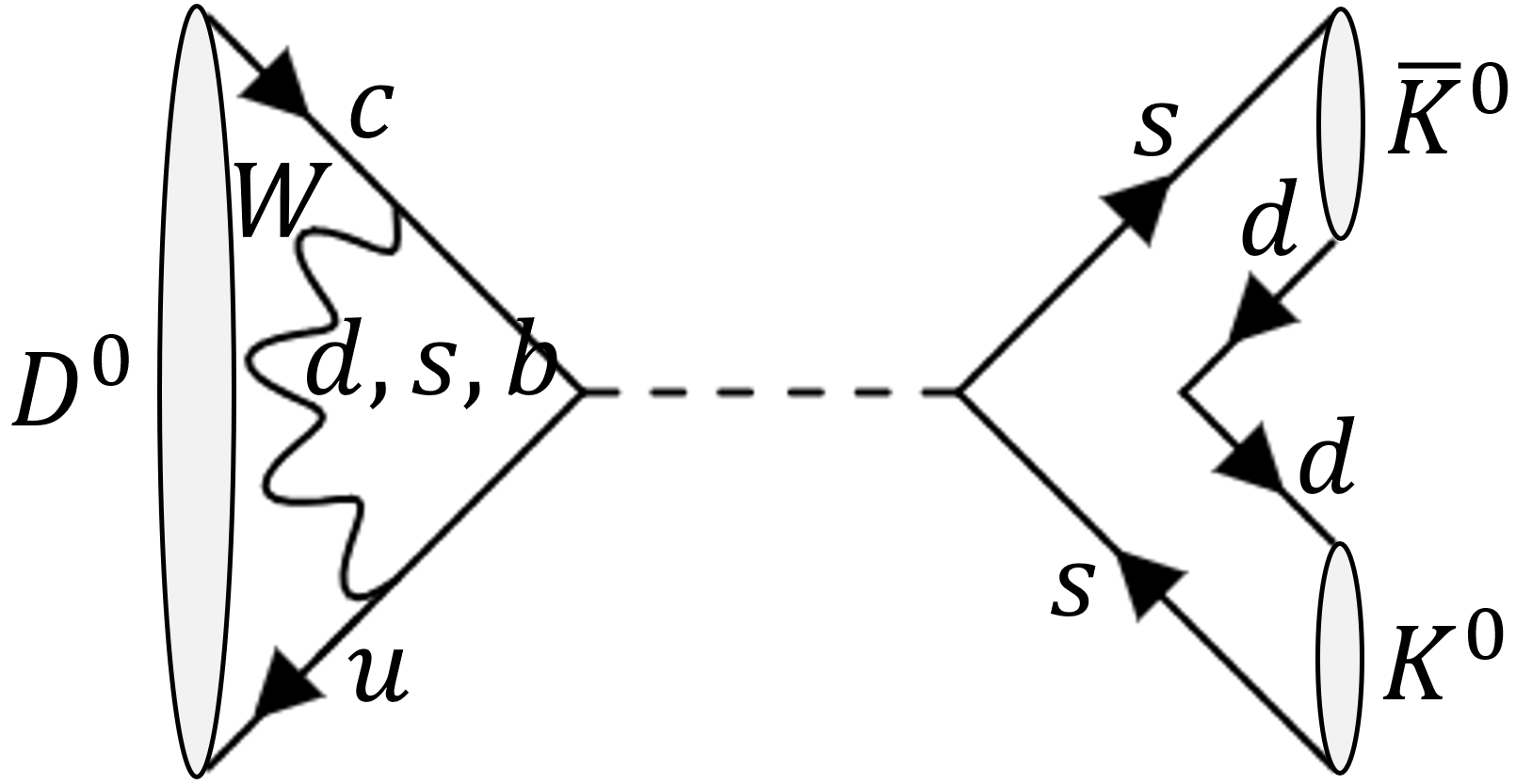}
    \caption{Diagrammatic representation of the topologies contributing to the $D^0\to K^0 \bar {K}^0$ decay in the SM. The exchange topologies are distinguished by the internal quarks as $E_s$ (left) and $E_d$ (middle). The penguin annihilation $PA_q$ processes (right) proceed with $q=d,s,b$ quarks in the loop.}
    \label{fig:DtoK0K0bar Feynman diagrams}
\end{figure}

We now turn to the $D^0 \to K^0 \bar{K}^0$ decay, whose dynamics within the SM significantly differs from that of the charged final-state modes discussed above. In this channel, both tree and penguin topologies are absent and the decay can proceed only through non-factorisable contributions, namely exchange ($E$) and penguin-annihilation ($PA$) diagrams, as illustrated in Fig.~\ref{fig:DtoK0K0bar Feynman diagrams}. Consequently, the amplitude can be written as follows:
\begin{equation}\label{eq:k0k0}
    A(D^0\to K^0\bar{K}^0) = \lambda_s E_d + \lambda_d E_s + \sum_{q=d,s,b
}\lambda_q \; PA_q \ , 
\end{equation}
where the subscript indicates the internal $q \bar{q}$ pair. As two distinct exchange amplitudes contribute to this decay, we distinguish them explicitly: $E_s$ denotes the exchange amplitude from the $c\to d$ transition with an internal $s\bar s$ pair while $E_d$ corresponds to the $c\to s$ transition with an internal $d\bar d$ pair. This structure differs from the
$D^0 \to \pi^+\pi^-$ and $D^0\to K^-K^+$ channels, where the exchange amplitudes $E$ and $E^\prime$ receive contribution solely from internal $u \bar u$ pairs, cf. Fig.~\ref{fig:DtoKK}. 

Using again the unitarity of the CKM matrix, cf.\ Eq.~\eqref{eq:ckmuni}, to eliminate $\lambda_d$, we write in analogy to Eq.~\eqref{eq:AmplKK}:
\begin{equation}\label{eq:AK0K0formula}
     A(D^0\to K^0\bar{K}^0) = \lambda_s \; \mathcal{A}_{K\bar{K}} \left[1+ \frac{\lambda_b}{\lambda_s} a_{K\bar{K}} e^{i\theta_{K\bar{K}}}  \right] \,,
\end{equation}
where
\begin{equation}
    \mathcal{A}_{K\bar{K}} = (E_d - E_s) + PA_{sd} \, ,   \qquad a_{K\bar{K}} \; e^{i\theta_{K\bar{K}}} \equiv \frac{- E_s + PA_{bd}}{(E_d - E_s) + PA_{sd}}\,,
\end{equation}
and $PA_{qr}\equiv PA_q - PA_r$ denotes the difference of penguin-annihilation diagrams with internal $q$ and $r$ quarks. 

The CP-averaged rate is given as in Eq.~\eqref{eq:CP-averaged-rate} by replacing $KK\to K\bar{K}$, leading to the following expression for the branching ratio:
\begin{equation}
{\cal B}(D^0 \to K^0\bar{K}^0) = {\cal N}_{K\bar{K}} |\lambda_s|^2|\mathcal{A}_{K\bar{K}}|^2 \tau(D^0) \left[1+ 2 \left|\frac{\lambda_b}{\lambda_s}\right|\; a_{K\bar{K}}\cos\theta_{K\bar{K}} \cos\gamma +  \left|\frac{\lambda_b}{\lambda_s}\right|^2 a_{K\bar{K}}^{2}\right]  \,,
\label{eq:BRKKbar}
\end{equation}
where the phase-space factor ${\cal N}_{K \bar K}$ is given in Eq.~\eqref{eq:curlyN}. Defining $|K_{\rm S}^0 \rangle \equiv (|K^0\rangle - |\bar K^0\rangle)/\sqrt{2}$, which neglects tiny effects from CP violation in the kaon sector at the level of $10^{-3}$, the branching ratio for the  $D^0 \to K_{\rm S}^0 K^0_{\rm S}$ mode is obtained as
\begin{equation}
{\cal B}(D^0 \to K^0_{\rm S} K^0_{\rm S}) = \frac12 {\cal B}(D^0 \to K^0 \bar K^0)\,.
\end{equation}

In the $U$-spin limit, we have $E_s = E_d$ and similarly $PA_{sd}=0$. In this case, $\mathcal{A}_{K\bar{K}}$ vanishes and, therefore, the branching ratio would only be proportional to the CKM subleading terms $a_{K\bar{K}}$, suppressed by $|\lambda_b/\lambda_s|\sim \lambda^4 \sim 10^{-3}$. 
However, this decay is  experimentally well established, with a branching ratio, cf.\ Eq.~\eqref{eq:Br_KsKs_exp-0}, only about a factor of ten smaller than those of the other SCS decays $D^0\to \pi^- \pi^+$ and $D^0\to K^- K^+$ in Eqs.~\eqref{eq:Br_pipi_exp-0} and \eqref{eq:Br_KK_exp-0}.
This would thus suggest the presence of sizeable non-factorisable contributions as well as significant $U$-spin breaking effects. 

Motivated by this intriguing pattern and by our findings from the analysis of non-factorisable contributions in the charged final states discussed in Section~\ref{sec:nonfac}, in the following 
we investigate this mode within the SM to determine under which conditions the experimental data can be accommodated, and what this implies for the magnitude of non-factorisable contributions and $U$-spin-breaking effects in $D^0 \to K_{\rm S}^0 K_{\rm S}^0$.

 To proceed, we isolate in the leading CKM amplitude in Eq.~\eqref{eq:AK0K0formula} the non-factorisable contribution arising from the $c \to s$ transition, and write~\footnote{Since the $D^0 \to K^0 \bar{K}^0$ decay only contains non-factorisable contributions, we cannot factorise the leading amplitude as in the $D^0 \to K^+K^-$ case. }

\begin{equation}
{\cal A}_{K\bar{K}}  =  (E_d + PA_{s}) \left( 1 - r_{K \bar K} e^{i \delta_{K \bar K}} \right) \ 
\label{eq:AmplKK-decomposition}
\end{equation}
where 
\begin{equation}
r_{K \bar K} \; e^{i \delta_{K\bar{K}}} \equiv \frac{ E_s + PA_d}{E_d + PA_s} 
\label{eq:rKKbar_deltaKKbar}
\end{equation}
parametrises the size of $U$-spin breaking effects in the leading decay amplitude ${\cal A}_{K \bar K}$. In the exact $U$-spin limit, we would have $r_{K \bar K} = 1$ and $\delta_{K \bar K} =~0$.


Starting from the parametrisation of the decay amplitude given in Eq.~\eqref{eq:AmplKK-decomposition}, and neglecting tiny additional corrections proportional to $|\lambda_b/\lambda_s| \sim \lambda^4\sim 10^{-3}$, we write the branching ratio in Eq.~\eqref{eq:BRKKbar} as 
\begin{equation}
{\cal B}(D^0 \to K^0 \bar K^0) = {\cal N}_{K \bar K} |\lambda_s|^2 |(E_d + PA_{s})|^2 \tau(D^0) \Big[ 1  - 2 r_{K \bar K} \cos \delta_{K \bar K} + r_{K \bar K}^2\Big]\,.
\label{eq:Br_KKbar}
\end{equation}

\subsection{\boldmath Connecting to the $D^0\to K^-K^+$ decay}
In order to investigate what the $D^0\to K_{\rm S}^0 K_{\rm S}^0$ branching ratio measurement in Eq.~\eqref{eq:Br_KsKs_exp-0} implies for the structure of this decay within the SM and, in particular, for the size of the $U$-spin breaking effects, we make the following two assumptions:
\begin{itemize}
    \item[(i)] {\it We assume that in $D^0\to K^0 \bar{K}^0$ the size of the penguin-annihilation contributions $(PA)$ is suppressed compared to that of the exchange $(E)$ diagrams and play a minor role.} This assumption is motivated by naive power counting, as penguin-annihilation diagrams are loop-suppressed compared to the exchange contributions and require colour-singlet exchange, represented by the dashed line in Fig.~\ref{fig:DtoK0K0bar Feynman diagrams}. Under this assumption, $r_{K\bar{K}}$ probes the ratio $|E_s/E_d|$ which differs from one only through $U$-spin-breaking effects.  
    \item[(ii)]     {\it We assume that the non-factorisable effects in $D^0\to K^- K^+$ parametrised by $r_{KK}$ in Eq.~\eqref{eq:rf} are dominated by the exchange topology.}  
    We recall that 
    \begin{equation}
        r_{KK} = \left|\frac{E + P_{sd} + T_{KK}^{\rm non-fac}}{T_{KK}^{\rm fac}}\right| \ ,
    \end{equation}
    where the penguin topologies $P_{sd}$ are further suppressed, appearing only through the difference of loop-diagrams with internal strange and down quarks. Therefore, our approximation actually consists in assuming that non-factorisable corrections to the tree topology are smaller than the exchange topology.
\end{itemize}
Under the assumptions (i) and (ii), the non-factorisable contribution from the $c \to s$ transition in Eq.~\eqref{eq:Br_KKbar} can be extracted from the value of the parameter $r_{KK}$, since the corresponding exchange amplitudes differ only by interchanging $d \leftrightarrow u$, and are equal in the limit of isospin symmetry. Assuming isospin symmetry and the points listed above, we write
\begin{equation}
{\cal N}_{K \bar K} |\lambda_s|^2 |E_d + PA_s|^2 \tau (D^0) = {\cal B}(D^0 \to K^- K^+)|_{\rm fac}\,  r_{K K}^2\,,
\end{equation}
yielding
\begin{equation}
{\cal B}(D^0 \to K^0_{\rm S} K^0_{\rm S}) = \frac{1}{2} \Big[ 1 + r_{K \bar K}^2 - 2 r_{K \bar K} \cos \delta_{K \bar K} \Big] \,  {\cal B}(D^0 \to K^- K^+)|_{\rm fac} \, r_{K K}^2 \,,
\label{eq:Br_KsKs}
\end{equation}
where we use Eq.~\eqref{eq:Br_fac_FNAL} for the factorised branching ratio. 

Fitting Eq.~\eqref{eq:Br_KsKs} to the experimental result in Eq.~\eqref{eq:Br_KsKs_exp-0}, we study how the allowed parameter space for the $U$-spin breaking parameters $r_{K \bar K}$ and $\delta_{K \bar K}$ changes as function of the input value for $r_{KK}$. For this parameter, we use only values allowed by the $D^0\to K^-K^+$ branching ratio,~\footnote{Note that formally, by keeping the parameter $r_{KK}$ arbitrary, the parametrisation in Eq.~\eqref{eq:Br_KsKs} could be written without the need to make assumptions (i) and (ii). In this case, however, the size of $r_{KK}$ would be completely unconstrained.} derived in the previous section and given in the right panel of Fig.~\ref{fig:BrP1P2}. 

\begin{figure}[t]
\centering
\includegraphics[scale = 1]{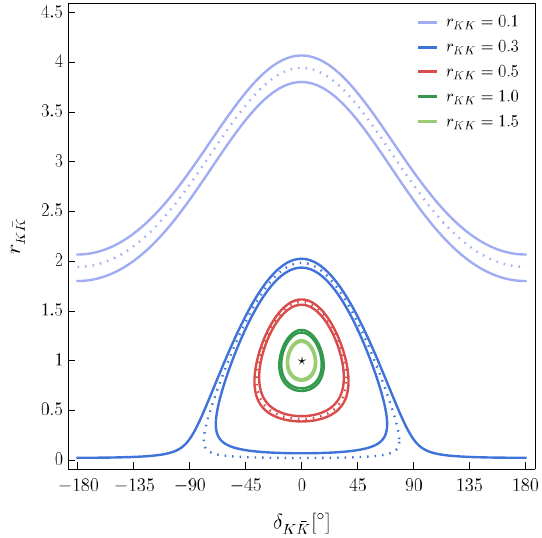}
\caption{Allowed $1\sigma$ ranges for the $U$-spin-breaking parameters $r_{K\bar K}$ and $\delta_{K \bar K}$, defined in Eq.~\eqref{eq:rKKbar_deltaKKbar}, obtained from Eq.~\eqref{eq:Br_KsKs}. 
The point representing the $U$-spin-symmetric case is indicated by a black star.}
\label{fig:BrKsKs}
\end{figure}
Specifically, we consider $r_{KK} = \{0.1, 0.3, 0.5, 1.0, 1.5\}$. The allowed $1\sigma$ regions for $r_{K\bar{K}}$ and its phase are shown in Fig.~\ref{fig:BrKsKs}. Here we only include the uncertainty on the factorised amplitude in Eq.~\eqref{eq:Br_fac_FNAL}. The experimental uncertainties coming from ${\cal B}(D^0 \to K^0_{\rm S} K^0_{\rm S})$ fall within this range. 
We see that as the value of $r_{KK}$ increases, the size of $U$-spin-breaking contributions needed to accommodate the experimental branching ratio reduces, as expected. Very small values of $r_{KK}$, e.g.\ $r_{KK} \sim 0.1$ would imply huge $U$-spin breaking effects of at least ${\cal O}(100\%)$. In contrast, large values of the exchange topology in $D^0 \to K^-K^+$, i.e.\ $r_{KK} \geq 1$ would strongly limit the allowed parameter space, constraining it to be very close to the $U$-spin symmetric point $(0,1)$. On the other hand, taking $r_{KK} \sim 0.5$ leads to the most plausible SM physics scenario, with $U$-spin-breaking effects of ${\cal O}(50\%)$. 


We conclude that, as for $D^0\to \pi^-\pi^+$ and $D^0\to K^-K^+$, we can accommodate the current experimental data with moderate $U$-spin breaking effects.

\subsection{Results for direct CP violation}
Finally, we study the implications that the obtained contour regions for the $U$-spin-breaking parameters $r_{K \bar K}$ and $\delta_{K \bar K}$ in Fig.~\ref{fig:BrKsKs} have on the expected size of the direct CP violation in the $D^0 \to K_{\rm S}^0 K_{\rm S}^0$ channel within the SM. The direct CP asymmetry is defined in Eq.~\eqref{eq:acpdef}. 

For this observable, the strongly suppressed $\lambda_b/\lambda_s$ terms paramaterised by $a_{K\bar{K}}$ and $\theta_{K\bar{K}}$ in Eq.~\eqref{eq:AK0K0formula} play the key role. 
The CP asymmetry can be readily obtained from Eq.~\eqref{eq:acpa} in terms of these parameters by replacing $KK \to K\bar{K}$, allowing for a quick comparison with the $D^0\to K^- K^+$ modes. However, for our purposes, it is beneficial to write the CP asymmetry in terms of the $U$-spin-breaking parameter $r_{K\bar{K}}$ and its phase. We define 
\begin{equation}
    a_{K\bar{K}}\; e^{i\theta_{K\bar{K}}} \equiv  r_{K\bar{K}}^p e^{i\delta_{K\bar{K}}^p} - r_{K\bar{K}} e^{i\delta_{K\bar{K}}} \ ,
\end{equation}
where
\begin{equation}
    r_{K\bar{K}}^p \; e^{i\delta_{K\bar{K}}^p }\equiv \frac{PA_b}{E_d + PA_s}
\end{equation}
denotes the ratio of the contribution to the amplitude due to the penguin operators and the current--current operators with external strange quarks.

For completeness, we give the total decay amplitude in Eq.~\eqref{eq:AK0K0formula} including the suppressed terms, which reads
\begin{equation}\label{eq:AK0K0tot}
     A(D^0\to K^0\bar{K}^0) = \lambda_s \; (E_d + PA_s)\left[1- r_{K\bar{K}} e^{i \delta_{K\bar{K}}} + \frac{\lambda_b}{\lambda_s} \left(  r_{K\bar{K}}^p e^{i\delta_{K\bar{K}}^p} - r_{K\bar{K}} e^{i\delta_{K\bar{K}}} \right) \right] \, .
\end{equation}
Using the above expressions, the direct CP asymmetry takes the following form:
\begin{align}
a_{\rm CP}^{\rm dir} (K^0_{\rm S} K^0_{\rm S}) =   2 {}&\left| \frac{\lambda_b}{\lambda_s} \right| \sin \gamma \left| \frac{r^p_{K \bar K} e^{i \delta^p_{K \bar K}} -  r_{K \bar K} e^{i \delta_{K \bar K}}  }{1- r_{K \bar K} e^{i \delta_{K \bar K}} } \right| \nonumber \\
{}& \times \sin \left[ \arg \left( \frac{ r^p_{K \bar K} e^{i \delta^p_{K \bar K}} - r_{K \bar K} e^{i \delta_{K \bar K}}  }{1- r_{K \bar K} e^{i \delta_{K \bar K}} }\right) \right] + \mathcal{O}\left(\left|\frac{\lambda_b}{\lambda_s}\right|^2\right)\,.
\label{eq:acp_KsKs}
\end{align}
Note that this observable would vanish in the $U$-spin limit.

\begin{figure}[t]
    \centering
   \includegraphics[width=0.7\linewidth]{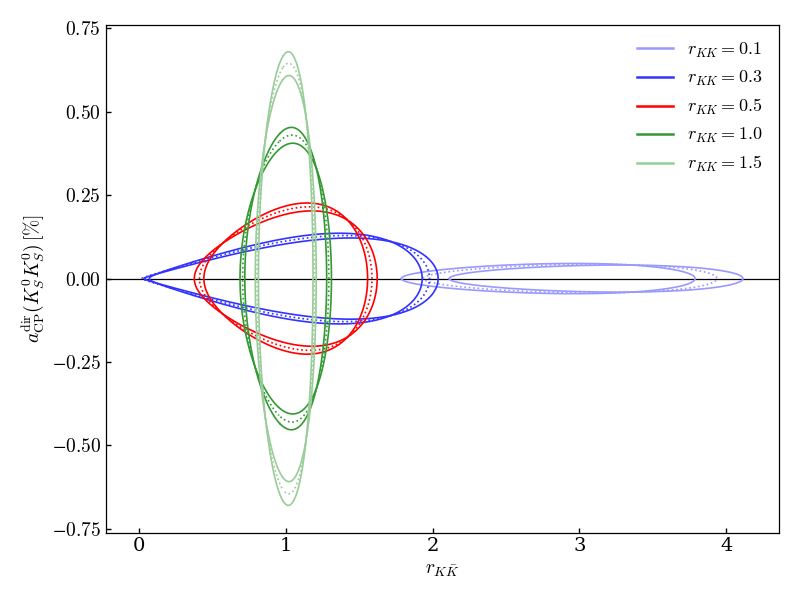}
    \caption{Direct CP asymmetry in $D^0 \to K_{\rm S}^0 K_{\rm S}^0$ as a function of the amplitude ratio $r_{K\bar{K}}$ for fixed values of $r_{KK}$. For each choice of $r_{KK}$, the CP asymmetry is computed along the central (dotted) and $1\sigma$ (solid) curves in Fig.~\ref{fig:BrKsKs} using Eq.~\eqref{eq:acp_KsKs}. Note that in this plot, the contribution of the penguin parameters $r_{K\bar{K}}^p$ and $\delta_{K\bar{K}}^p$ is set to zero.}
    \label{fig:aCP_KsKs_combi}
\end{figure}

\begin{figure}[h!]
    \centering
\includegraphics[width=0.7\linewidth]{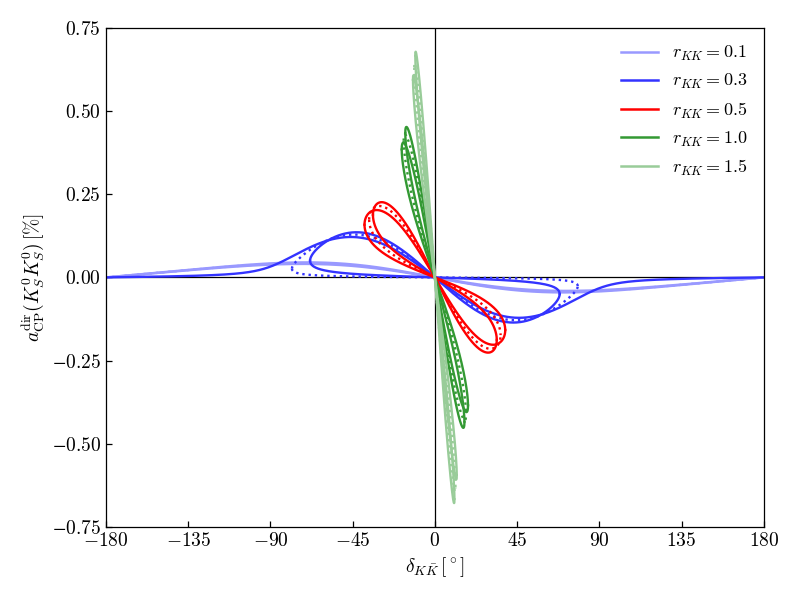}
    \caption{Direct CP asymmetry in $D^0 \to K_{\rm S}^0 K_{\rm S}^0$ as a function of the relative strong phase $\delta_{K\bar{K}}$ for fixed values of $r_{KK}$. For each choice of $r_{KK}$, the CP asymmetry is computed along the central (dotted) and $1\sigma$ (solid) curves in Fig.~\ref{fig:BrKsKs} using Eq.~\eqref{eq:acp_KsKs}. Note that in this plot, the contribution from the penguin parameters $r_{K\bar{K}}^p$ and $\delta_{K\bar{K}}^p$ is set to zero.}
    \label{fig:aCP_KsKs_combimag}
\end{figure}

Even using the constraints on the size of the $U$-spin-breaking parameters $(\delta_{K \bar K}, r_{K \bar K})$ obtained from the different contours in Fig.~\ref{fig:BrKsKs}, we cannot compute the CP asymmetry in Eq.~\eqref{eq:acp_KsKs}, as this also depends on the size of $r_{K \bar K}^p$, $\delta_{K \bar K}^p$ originating from the penguin-annihilation amplitudes. This contribution shown in Fig.~\ref{fig:DtoK0K0bar Feynman diagrams} involves a colour-singlet exchange (given by the dashed line) and is expected to have a minor impact on the CP asymmetry. As such, we first consider the case $r_{K \bar K}^p = 0$, $\delta_{K \bar K}^p = 0$. Then, for each point $(\delta_{K \bar K}, r_{K \bar K})$ in Fig.~\ref{fig:BrKsKs} we can compute the CP asymmetry using Eq.~\eqref{eq:acp_KsKs}. The results for the reference values $r_{KK} = \{0.1, 0.3, 0.5, 1.0, 1.5\}$ are given in Figs.~\ref{fig:aCP_KsKs_combi} and~\ref{fig:aCP_KsKs_combimag} as functions of $r_{K\bar{K}}$ and the strong phase $\delta_{K\bar{K}}$, respectively. We observe that the predicted direct CP asymmetry increases for values of $r_{K\bar{K}}$ closer to the $U$-spin limit, $r_{K\bar{K}}=1$ and $\delta_{K\bar{K}}=0$. 

\begin{figure}
\centering
\includegraphics[scale = 0.8]{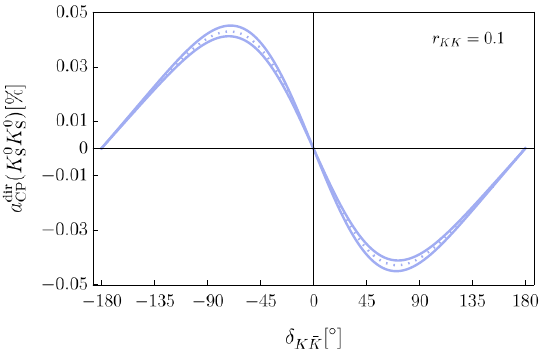} \qquad
\includegraphics[scale = 0.8]{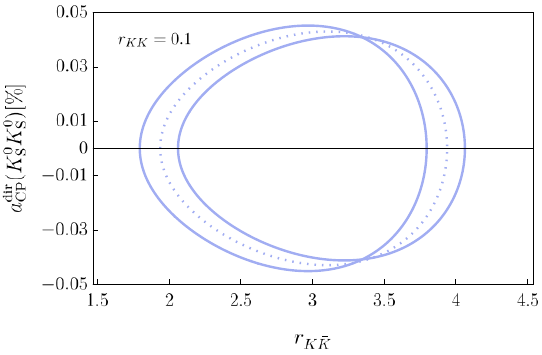}\\[4mm] 
\includegraphics[scale = 0.8]{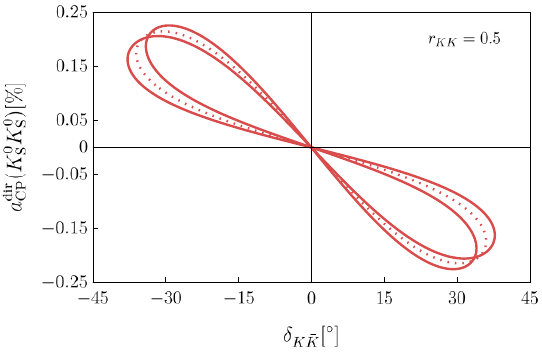} \qquad
\includegraphics[scale = 0.8]{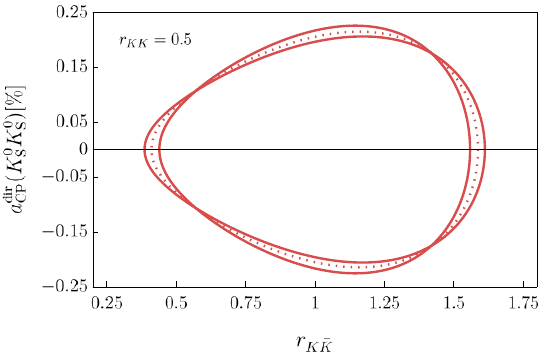}\\[4mm] 
\includegraphics[scale = 0.8]{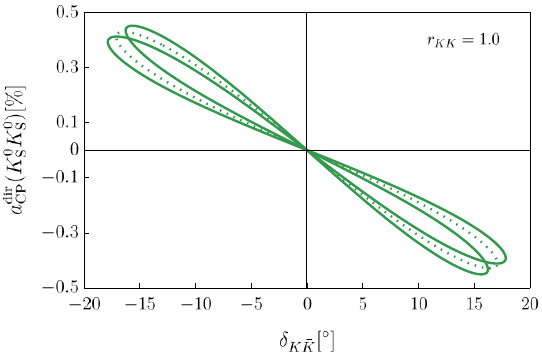} \qquad
\includegraphics[scale = 0.8]{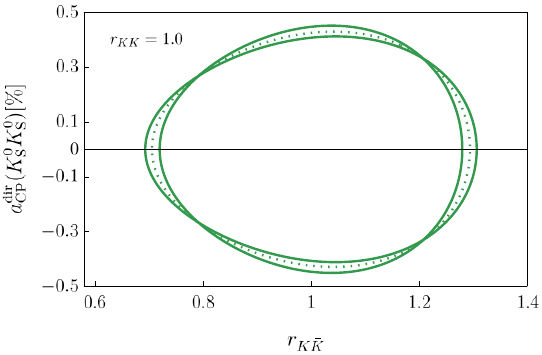}
\caption{The direct CP asymmetry $a_{\rm CP}^{\rm dir}$ given in Eq.~\ref{eq:acp_KsKs} as a function of the $U$-spin breaking parameters $\delta_{K \bar K}$ (left panels) and $r_{K \bar K}$ (right panels). The contribution of the penguin-annihilation parameter $r_{K\bar{K}}^p$ is set to zero.}
\label{fig:acpKsKs}
\end{figure}

In our benchmark scenario with $r_{KK} = 0.5$ and $U$-spin breaking effects of about 50$\%$, we find an upper bound for the direct CP asymmetry in $D^0 \to K_{\rm S}^0 K_{\rm S}^0$ of a few permille. 

For $r_{KK}=1.5$, the direct CP asymmetry could reach about a percent for very moderate $U$-spin-breaking contributions of about 10$\%$.  However, this would require very large exchange contributions in $D^0 \to K^- K^+$ of about $150\%$ with respect to the colour-allowed tree amplitude. 
Conversely, for $r_{KK} = 0.1$, only very large values of $U$-spin-breaking effects above 100$\%$ are allowed. We include this case for illustration and find a prediction for the direct CP asymmetry at the sub-permille level. 

In Fig.~\ref{fig:acpKsKs}, we show the direct CP asymmetries for the representative choices $r_{KK}=\{0.1, 0.5, 1.0\}$ separately, which allows us to illustrate more clearly the range and maximal size of the predicted effects. 

\begin{figure}
\centering
\includegraphics[width=0.8\linewidth]{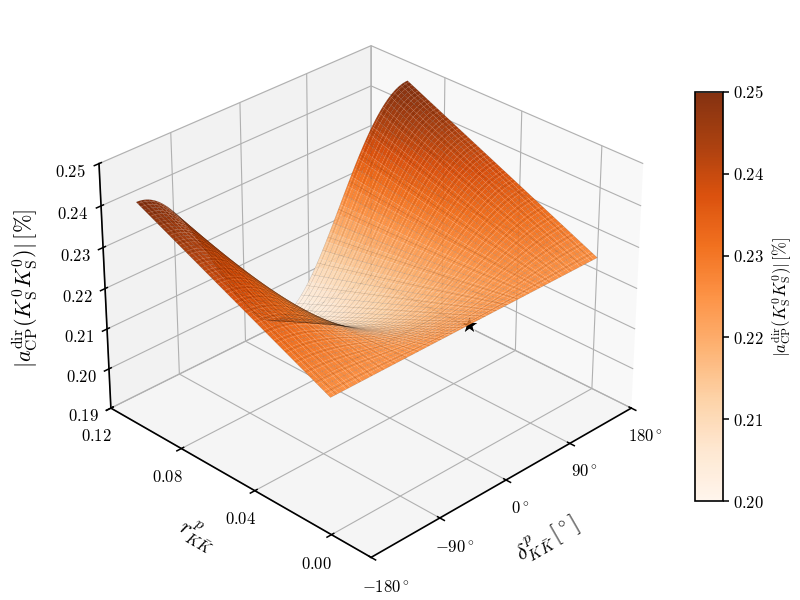}
\caption{Three-dimensional map of the direct CP asymmetry $|a_{\rm CP}^{\rm dir}(K_{\rm S}^0 K_{\rm S}^0)|$ as a function of the penguin magnitude $r_{K\bar{K}}^p$ and phase $\delta_{K\bar{K}}^p$ for the benchmark scenario $r_{KK} = 0.5$, where each point on the surface corresponds to the maximally allowed CP asymmetry. The black star marks the case of $r_{K \bar K}^p = \delta_{K \bar K}^p = 0$, where there is no penguin contribution.} 
\label{fig:acpKsKs with penguins}
\end{figure}

Finally, we also investigate how our predictions change when allowing for non-zero values of $r^p_{K\bar{K}}$ and $\delta^p_{K\bar{K}}$. 
To this end, we vary $r_{K \bar K}^p \in [0, 0.1]$, taking into account the expected suppression of the penguin-annihilation contribution, as well as $\delta_{K \bar K}^p \in [-  180^\circ, 180 ^\circ]$. 
In Fig.~\ref{fig:acpKsKs with penguins}, we show the  
resulting maximal size of the CP asymmetry for fixed $r_{KK} = 0.5$. The point marked with the black star corresponds to the case of $r_{K \bar K}^p = \delta_{K \bar K}^p = 0$, where the penguin contribution is neglected. This point represents the maximum of the red curves in Fig.~\ref{fig:acpKsKs}.
We find that allowing for a non-zero penguin contribution within the
range considered modifies the asymmetry only at the $10^{-3}$ level, indicating that neglecting these corrections has a minimal impact on the predicted direct CP asymmetry. While we only illustrate the result for $r_{KK}=0.5$, the conclusion remains unchanged for other values of $r_{KK}$: penguin-annihilation contributions up to $10\%$ have only a minimal impact on the maximum asymmetry obtained.

In order to compare with the experimental determinations, we note that the current world-average for the time-integrated CP asymmetry of $D^0\to K_{\rm S}^0K_{\rm S}^0$ given in Eq.~\eqref{eq:ACPKSKS} can be related to $a_{\rm CP}^{\rm dir}$. This requires properly including the time evolution due to $D^0$--$\bar D^0$ mixing (see e.g., \cite{Nierste:2015zra}). Given the current experimental precision, these effects can be neglected and $A_{\rm CP}(K_{\rm S}^0 K_{\rm S}^0)|_{\rm exp}$ effectively corresponds to direct CP violation. However, once the experimental precision increases, time-dependent effects will need to be properly included. 
At the per-mille level, also effects due to CP violation in the neutral $K$ system will need to be taken into account (for a discussion see~\cite{Grossman:2011zk}).

With these caveats in mind, the current level of precision of the experimental world-average in Eq.~\eqref{eq:ACPKSKS} is at the level of the $\%$-regime. Our benchmark thus gives CP asymmetries at a significantly lower level.



\section{Conclusions and Outlook}\label{sec:concl}

\begin{figure}[t]
    \centering
    \includegraphics[width=0.8\linewidth]{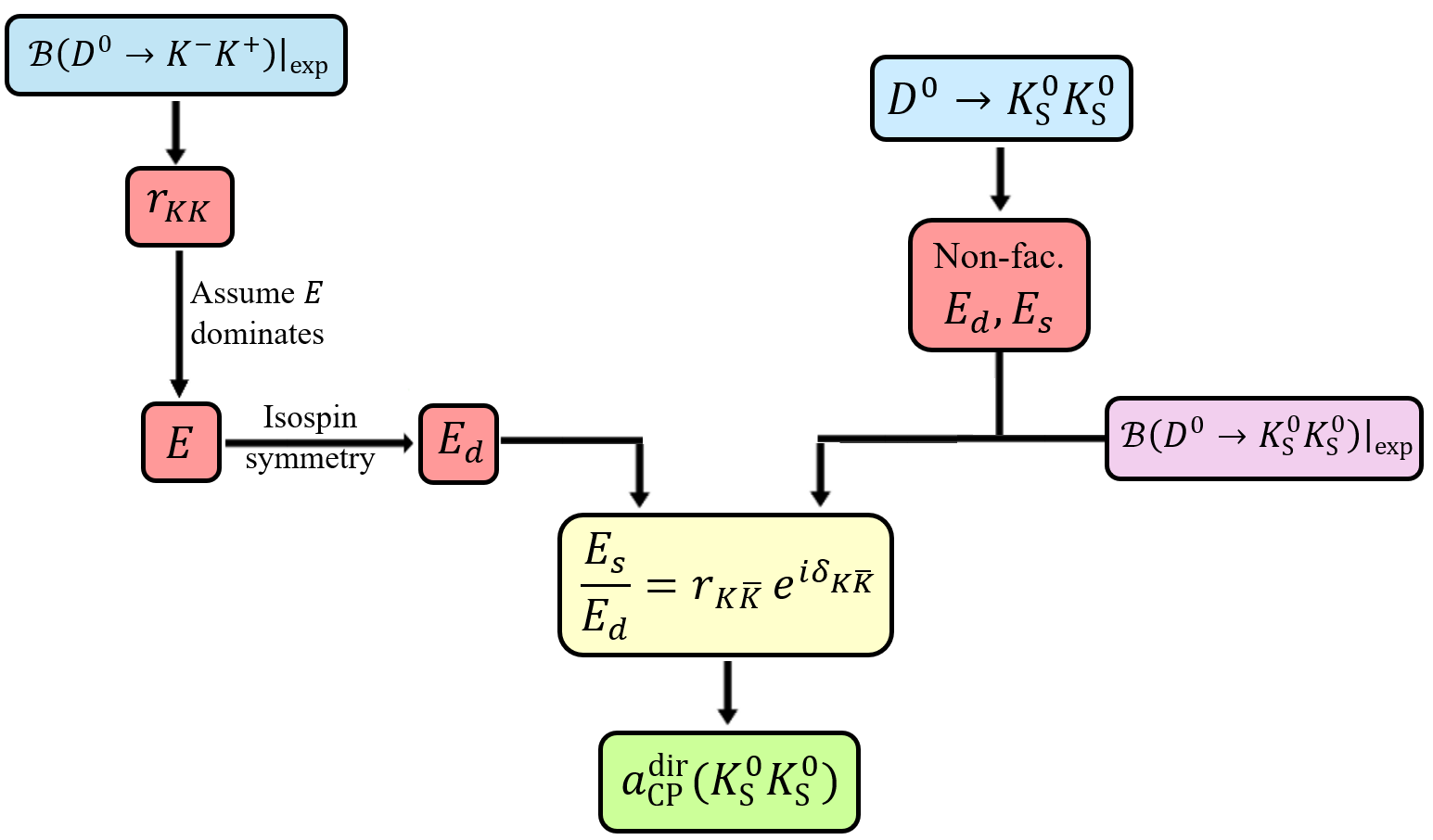}
    \caption{Overview of our strategy to study SM benchmarks for the $D^0\to K_{\rm S}^0K_{\rm S}^0$ decay.} 
    \label{fig:Kshort flowchart}
\end{figure}
We have performed a detailed analysis of the non-leptonic $D^0\to K^- K^+, \pi^-\pi^+$ and $K^0_{\rm S} K^0_{\rm S}$ decays within 
the SM. Employing results for non-perturbative form factors and decay constants from lattice QCD, we find that 
the factorisation of the hadronic matrix elements of the current--current operators into hadronic matrix elements of quark currents  results 
naturally in the pattern of the measured branching ratios of the $D^0\to K^- K^+$ and $D^0\to \pi^-\pi^+$ channels governed by colour-allowed tree topologies, while still leaving space for non-factorisable and $U$-spin-breaking contributions at the 50\% level. Although these effects are significantly larger than in non-leptonic $B$-meson decays with similar dynamics, in view of the smaller charm-quark mass in comparison with $m_b\sim 5$\,GeV, such effects are falling well into the ballpark of the theoretical expectations. Concerning experimental results for direct CP violation in the $D^0\to K^- K^+, \pi^-\pi^+$ system, we find that the required minimum value of the ratio of penguin to colour-allowed tree amplitudes is $1.6 \pm 0.5$ and $0.7 \pm 0.6$ for the $D^0\to \pi^-\pi^+$ and $D^0\to K^- K^+$ decays, respectively, indicating a significant enhancement of the penguin contributions. It is also well possible that CP-violating New Physics contributions enter at a level too small to significantly affect the branching ratios, for example through new particles in flavour-changing neutral current penguin loops, resulting in the pattern of the measured direct CP asymmetries. The current uncertainties are too large to draw further conclusions. 

A fascinating probe for non-factorisable strong interaction dynamics is provided by the $D^0\to K^0_{\rm S} K^0_{\rm S}$ mode which originates only from exchange and penguin annihilation topologies in the SM. In the exact $U$-spin limit, the corresponding decay amplitude would vanish up to a negligible remnant which is strongly CKM suppressed by ${\cal O}(10^{-3})$ with respect to the leading terms. Nevertheless, the decay is experimentally well established with a branching ratio measured with a precision at the $5\%$ level. We addressed the question of how to accommodate this result in the SM. Assuming that the non-factorizable contribution of the $D^0\to K^- K^+$ decay is dominated by the exchange topology, we use the isospin symmetry of the strong interactions to determine the counterpart in $D^0\to K^0_{\rm S} K^0_{\rm S}$. The other exchange amplitude can then be determined through the measured branching $D^0\to K^0_{\rm S} K^0_{\rm S}$ ratio. Our strategy is illustrated in the flowchart in Fig.~\ref{fig:Kshort flowchart}. Following this method, we find that again $U$-spin-breaking effects in the exchange amplitudes at the 
50\% level allow us to accommodate the data in the SM. Finally, utilising these results, we performed a study of the direct CP asymmetry in $D^0\to K^0_{\rm S} K^0_{\rm S}$, finding upper bounds within our SM benchmarks at the few per-mille level. 

The current experimental world average is at the percent level, thereby leaving a lot of space with respect to these ranges. It is well possible that more precise measurements of this CP asymmetry will lead to surprises, which may result from New Physics contributions entering with new sources of CP violation at a level too small to significantly affect the branching ratio. At the end of Belle II, a sensitivity of $15\times 10^{-4}$ may be reached for the time-integrated CP asymmetry $A_{\rm CP}$ of 
$D^0\to K^0_{\rm S} K^0_{\rm S}$, while the FCC-ee may even achieve $2 \times 10^{-4}$, which would be about 50 
times better than the current measurement~\cite{FCC}. These exciting prospects will certainly keep both the theoretical and the experimental communities busy in the years ahead of us.

\section*{Acknowledgements}
This research has been supported by the Dutch Research Council (NWO). The work of K.K.V. is supported in part by the Dutch Research Council (NWO) as part
of the project Solving Beautiful Puzzles (VI.Vidi.223.083) of the research programme Vidi.


\appendix
\section{Effective operators and Wilson coefficients}
\label{app:1}
In order to factorise the amplitude, we  describe the $D^0 \to \pi^- \pi^+$ and $D^- \to K^- K^+$ within 
the framework of the low-energy effective Hamiltonian. 
At the renormalisation scale $\mu \sim m_c$, the Hamiltonian governing the underlying flavour-changing non-leptonic charm-quark decays $c \to q \bar q u$ with $q = \{u,d,s\}$ is given as follows~\cite{Buchalla:1995vs}:
\begin{equation}
{\cal H}_{\rm eff} = \frac{G_F}{\sqrt{2}} \Big[ \sum_{q = d,s} \lambda_{q}\left(C_1 Q_1^{q} + C_2 Q_2^{q}\right) - \lambda_b \sum_{k = 3}^6 C_k Q_k \Big] + {\rm h.c.} \,,
\label{eq:Eff_Ham}
\end{equation}
where the current--current operators $Q_1^q$ and $Q_2^q$ are defined as~\cite{Buchalla:1995vs}
\begin{align}
Q_1^q = \left[ \bar q^i \gamma_\mu(1-\gamma_5) c^j \right]\left[\bar u^j \gamma^\mu(1 - \gamma_5) q^i \right]\,,
\label{eq:Q1}
\\[2mm]
Q_2^q = \left[ \bar q^i \gamma_\mu(1-\gamma_5) c^i \right]\left[ \bar u^j \gamma^\mu(1- \gamma_5) q^j \right]\,,
\label{eq:Q2}
\end{align}
where $i,j$ are colour indices. Additionally, the QCD penguin operators $Q_k$, with $k=1,\ldots, 6$ read~\footnote{Note that the contribution of additional operators such as the electroweak penguins can be safely neglected for the analysis of the decays considered in our study.} 
\begin{align}
Q_3 = \left[ \bar u^i \gamma_\mu(1-\gamma_5) c^i \right] \!\sum_{q = u,d,s}\!\left[ \bar q^j \gamma^\mu(1-\gamma_5) q^j \right]\,,
\\[2mm]
Q_4 = \left[ \bar u^i \gamma_\mu(1-\gamma_5) c^j \right] \!\sum_{q = u,d,s} \!\left[ \bar q^j \gamma^\mu(1- \gamma_5) q^i \right]\,,
\label{eq:Q4}
\\[2mm]
Q_5 = \left[ \bar u^i \gamma_\mu(1-\gamma_5) c^i \right] \!\sum_{q = u,d,s}\!\left[ \bar q^j \gamma^\mu(1+\gamma_5) q^j\right]\,,
\\[2mm]
Q_6 = \left[ \bar u^i \gamma_\mu(1-\gamma_5) c^j \right] \!\sum_{q = u,d,s}\! \left[ \bar q^j \gamma^\mu(1+\gamma_5) q^i\right] \,.
\label{eq:Q6}
\end{align}

The Wilson coefficients $C_k(\mu)$ in Eq.~\eqref{eq:Eff_Ham} are known to the NNLO-QCD accuracy~\cite{deBoer:2016dcg}. A comparison of their LO (NLO) results for the corresponding for different values of the renormalisation scale $\mu$ is shown in Table~\ref{tab:WC}.

\begin{table}[t]
    \centering
    \begin{tabular}{|c|c|c|c|c|c|c|}
    \hline
      $\mu\,$[GeV] & $C_1(\mu)$ & $C_2 (\mu)$ & $C_3(\mu)$ & $C_4(\mu)$ & $C_5(\mu)$ & $C_6(\mu)$ 
      \\
      \hline
      \hline
    1 & -0.64 (-0.50) & 1.35 (1.26) & 0.02 (0.03) & -0.04 (-0.06) &  0.01 (0.01) & -0.05 (-0.07)
    \\
    \hline
    1.5   & -0.50 (-0.37) & 1.24 (1.18) & 0.01 (0.01) & -0.02 (-0.04) & 0.01 (0.01) & -0.02 (-0.04)
    \\
    \hline
    3   & -0.32 (-0.24) & 1.15 (1.10) & 0.00 (0.00) & -0.01 (-0.01) & 0.00 (0.00) & -0.01 (-0.01)
    \\
    \hline
    \end{tabular}
    \caption{Comparison of the LO (NLO) values of the Wilson coefficients for charm-quark decays for different choices of the renormalisation scale $\mu$~\cite{Buchalla:1995vs}. The NLO results are obtained in the naive dimensional regularisation scheme.}
    \label{tab:WC}
\end{table}


\bibliographystyle{JHEP}
\bibliography{References}

@article{DeBruyn:2022zhw,
    author = "De Bruyn, Kristof and Fleischer, Robert and Malami, Eleftheria and van Vliet, Philine",
    title = "{New physics in $B_{q}^{0}$-mixing: present challenges, prospects, and implications for}",
    eprint = "2208.14910",
    archivePrefix = "arXiv",
    primaryClass = "hep-ph",
    reportNumber = "Nikhef-2022-012",
    doi = "10.1088/1361-6471/acab1d",
    journal = "J. Phys. G",
    volume = "50",
    number = "4",
    pages = "045003",
    year = "2023"
}

@article{LHCb:2022lry,
    author = "Aaij, R. and others",
    collaboration = "LHCb",
    title = "{Measurement of the Time-Integrated CP Asymmetry in $D^0 \to K^+K^-$ Decays}",
    eprint = "2209.03179",
    archivePrefix = "arXiv",
    primaryClass = "hep-ex",
    reportNumber = "CERN-EP-2022-163, LHCb-PAPER-2022-024",
    doi = "10.1103/PhysRevLett.131.091802",
    journal = "Phys. Rev. Lett.",
    volume = "131",
    number = "9",
    pages = "091802",
    year = "2023"
}

@article{DeBruyn:2012wk,
    author = "De Bruyn, Kristof and Fleischer, Robert and Knegjens, Robert and Koppenburg, Patrick and Merk, Marcel and Pellegrino, Antonio and Tuning, Niels",
    title = "{Probing New Physics via the $B^0_s\to \mu^+\mu^-$ Effective Lifetime}",
    eprint = "1204.1737",
    archivePrefix = "arXiv",
    primaryClass = "hep-ph",
    reportNumber = "NIKHEF-2012-006",
    doi = "10.1103/PhysRevLett.109.041801",
    journal = "Phys. Rev. Lett.",
    volume = "109",
    pages = "041801",
    year = "2012"
}

@article{Khodjamirian:2009ys,
    author = "Khodjamirian, A. and Klein, Ch. and Mannel, Th. and Offen, N.",
    title = "{Semileptonic charm decays $D \to \pi \ell \nu_\ell$ and $D \to K \ell \nu_\ell$
 from QCD Light-Cone Sum Rules}",
    eprint = "0907.2842",
    archivePrefix = "arXiv",
    primaryClass = "hep-ph",
    reportNumber = "SI-HEP-2009-06, LPT-09-60",
    doi = "10.1103/PhysRevD.80.114005",
    journal = "Phys. Rev. D",
    volume = "80",
    pages = "114005",
    year = "2009"
}

@article{Fleischer:2022klb,
    author = "Fleischer, Robert and Malami, Eleftheria and Rehult, Anders and Vos, K. Keri",
    title ="{Fingerprinting CP-violating New Physics with $B \to K \mu^+ \mu^-$}",
    eprint = "2212.09575",
    archivePrefix = "arXiv",
    primaryClass = "hep-ph",
    reportNumber = "Nikhef-2022-022, SI-HEP-2022-36, P3H-22-123",
    doi = "10.1007/JHEP03(2023)113",
    journal = "JHEP",
    volume = "03",
    pages = "113",
    year = "2023"
}

@article{Gambino:2020jvv,
    author = "Gambino, P. and others",
    title = "{Challenges in semileptonic $B$ decays}",
    eprint = "2006.07287",
    archivePrefix = "arXiv",
    primaryClass = "hep-ph",
    reportNumber = "FERMILAB-PUB-20-235-T",
    doi = "10.1140/epjc/s10052-020-08490-x",
    journal = "Eur. Phys. J. C",
    volume = "80",
    number = "10",
    pages = "966",
    year = "2020"
}

@article{Bernlochner:2024sfg,
    author = "Bernlochner, Florian U. and Prim, Markus T. and Vos, K. Keri",
    title = "{$|V_{ub}|$ and $|V_{cb}|$ from exclusive semileptonic decays}",
    doi = "10.1140/epjs/s11734-023-01077-z",
    journal = "Eur. Phys. J. ST",
    volume = "233",
    number = "2",
    pages = "347--358",
    year = "2024",
    note = "[Erratum: Eur.Phys.J.ST 233, 5--8 (2024)]"
}

@article{DeBruyn:2025rhk,
    author = "De Bruyn, Kristof and Fleischer, Robert and Malami, Eleftheria",
    title = "{How to tame penguins: Advancing to high-precision measurements of $\phi_d$ and $\phi_s$}",
    eprint = "2505.06102",
    archivePrefix = "arXiv",
    primaryClass = "hep-ph",
    reportNumber = "Nikhef-2025-007, SI-HEP-2025-09, P3H-25-031",
    month = "5",
    year = "2025"
}

@article{Charles:2004jd,
    author = "Charles, J. and Hocker, Andreas and Lacker, H. and Laplace, S. and Le Diberder, F. R. and Malcles, J. and Ocariz, J. and Pivk, M. and Roos, L.",
    collaboration = "CKMfitter Group",
    title = "{CP violation and the CKM matrix: Assessing the impact of the asymmetric $B$ factories}",
    eprint = "hep-ph/0406184",
    archivePrefix = "arXiv",
    reportNumber = "CPT-2004-P-030, LAL-04-21, LAPP-EXP-2004-01, LPNHE-2004-01",
    doi = "10.1140/epjc/s2005-02169-1",
    journal = "Eur. Phys. J. C",
    volume = "41",
    number = "1",
    pages = "1--131",
    year = "2005"
}

@article{ParticleDataGroup:2024cfk,
    author = "Navas, S. and others",
    collaboration = "Particle Data Group",
    title = "{Review of particle physics}",
    doi = "10.1103/PhysRevD.110.030001",
    journal = "Phys. Rev. D",
    volume = "110",
    number = "3",
    pages = "030001",
    year = "2024"
}

@article{FermilabLattice:2022gku,
    author = "Bazavov, Alexei and others",
    collaboration = "Fermilab Lattice, MILC",
    title = "{D-meson semileptonic decays to pseudoscalars from four-flavor lattice QCD}",
    eprint = "2212.12648",
    archivePrefix = "arXiv",
    primaryClass = "hep-lat",
    reportNumber = "MIT-CTP/5513, FERMILAB-PUB-22-943-T",
    doi = "10.1103/PhysRevD.107.094516",
    journal = "Phys. Rev. D",
    volume = "107",
    number = "9",
    pages = "094516",
    year = "2023"
}

@article{Lubicz:2017syv,
    author = "Lubicz, V. and Riggio, L. and Salerno, G. and Simula, S. and Tarantino, C.",
    collaboration = "ETM",
    title = "{Scalar and vector form factors of $D \to \pi(K) \ell \nu$ decays with $N_f=2+1+1$ twisted fermions}",
    eprint = "1706.03017",
    archivePrefix = "arXiv",
    primaryClass = "hep-lat",
    reportNumber = "PREPRINT-RM3-TH-17-6, preprint RM3-TH/17-6",
    doi = "10.1103/PhysRevD.96.054514",
    journal = "Phys. Rev. D",
    volume = "96",
    number = "5",
    pages = "054514",
    year = "2017",
    note = "[Erratum: Phys.Rev.D 99, 099902 (2019), Erratum: Phys.Rev.D 100, 079901 (2019)]"  
}

@article{Buchalla:1995vs,
    author = "Buchalla, Gerhard and Buras, Andrzej J. and Lautenbacher, Markus E.",
    title = "{Weak decays beyond leading logarithms}",
    eprint = "hep-ph/9512380",
    archivePrefix = "arXiv",
    reportNumber = "SLAC-PUB-7009, SLAC-PUB-95-7009, MPI-PH-95-104, TUM-T31-100-95, FERMILAB-PUB-95-305-T",
    doi = "10.1103/RevModPhys.68.1125",
    journal = "Rev. Mod. Phys.",
    volume = "68",
    pages = "1125--1144",
    year = "1996"
}

@article{deBoer:2016dcg,
    author = {de Boer, Stefan and M{\"u}ller, Bastian and Seidel, Dirk},
    title = "{Higher-order Wilson coefficients for $c \to u$ transitions in the standard model}",
    eprint = "1606.05521",
    archivePrefix = "arXiv",
    primaryClass = "hep-ph",
    reportNumber = "DO-TH-15-11, QFET-2015-27",
    doi = "10.1007/JHEP08(2016)091",
    journal = "JHEP",
    volume = "08",
    pages = "091",
    year = "2016"
}

@article{FLAG:2024oxs,
    author = "Aoki, Y. and others",
    collaboration = "Flavour Lattice Averaging Group (FLAG)",
    title = "{FLAG Review 2024}",
    eprint = "2411.04268",
    archivePrefix = "arXiv",
    primaryClass = "hep-lat",
    reportNumber = "CERN-TH-2024-192, FERMILAB-PUB-24-0785-T",
    month = "11",
    year = "2024"
}

@article{Lenz:2023rlq,
    author = "Lenz, Alexander and Piscopo, Maria Laura and Rusov, Aleksey V.",
    title = "{Two body non-leptonic D$^{0}$ decays from LCSR and implications for ${\Delta a}_{{\text{CP}}}^{{\text{dir}}}$}",
    eprint = "2312.13245",
    archivePrefix = "arXiv",
    primaryClass = "hep-ph",
    reportNumber = "SI-HEP-2023-34, P3H-23-105",
    doi = "10.1007/JHEP03(2024)151",
    journal = "JHEP",
    volume = "03",
    pages = "151",
    year = "2024"
}

@article{Wolfenstein:1983yz,
    author = "Wolfenstein, Lincoln",
    title = "{Parametrization of the Kobayashi-Maskawa Matrix}",
    reportNumber = "CMU-HEG83-9",
    doi = "10.1103/PhysRevLett.51.1945",
    journal = "Phys. Rev. Lett.",
    volume = "51",
    pages = "1945",
    year = "1983"
}

@article{Khodjamirian:2017fxg,
    author = "Khodjamirian, Alexander and Rusov, Aleksey V.",
    title = "{$B_{s}\to K \ell \nu_\ell$ and $B_{(s)} \to \pi (K) \ell^+\ell^-$ decays at large recoil and CKM matrix elements}",
    eprint = "1703.04765",
    archivePrefix = "arXiv",
    primaryClass = "hep-ph",
    reportNumber = "SI-HEP-2017-03, QFET-2017-03",
    doi = "10.1007/JHEP08(2017)112",
    journal = "JHEP",
    volume = "08",
    pages = "112",
    year = "2017"
}

@article{Fleischer:1999pa,
    author = "Fleischer, Robert",
    title = "{New strategies to extract Beta and gamma from $B_d \to \pi^+ \pi^-$ and $B_s \to K^+ K^-$}",
    eprint = "hep-ph/9903456",
    archivePrefix = "arXiv",
    reportNumber = "CERN-TH-99-79",
    doi = "10.1016/S0370-2693(99)00640-1",
    journal = "Phys. Lett. B",
    volume = "459",
    pages = "306--320",
    year = "1999"
}

@article{Fleischer:2022rkm,
    author = "Fleischer, Robert and Jaarsma, Ruben and Vos, K. Keri",
    title = "{Zooming into CP violation in $B_{(s)} \to hh$ decays}",
    eprint = "2211.08346",
    archivePrefix = "arXiv",
    primaryClass = "hep-ph",
    reportNumber = "Nikhef-2022-020",
    doi = "10.1007/JHEP02(2023)081",
    journal = "JHEP",
    volume = "02",
    pages = "081",
    year = "2023"
}

@article{LHCb:2019hro,
    author = "Aaij, Roel and others",
    collaboration = "LHCb",
    title = "{Observation of CP Violation in Charm Decays}",
    eprint = "1903.08726",
    archivePrefix = "arXiv",
    primaryClass = "hep-ex",
    reportNumber = "LHCb-PAPER-2019-006, CERN-EP-2019-042",
    doi = "10.1103/PhysRevLett.122.211803",
    journal = "Phys. Rev. Lett.",
    volume = "122",
    number = "21",
    pages = "211803",
    year = "2019"
}

@article{Khodjamirian:2017zdu,
    author = "Khodjamirian, Alexander and Petrov, Alexey A.",
    title = "{Direct CP asymmetry in $D\to \pi^-\pi^+$ and $D\to K^-K^+$ in QCD-based approach}",
    eprint = "1706.07780",
    archivePrefix = "arXiv",
    primaryClass = "hep-ph",
    reportNumber = "SI-HEP-2017-12, QFET-2017-09, WSU-HEP-1709",
    doi = "10.1016/j.physletb.2017.09.070",
    journal = "Phys. Lett. B",
    volume = "774",
    pages = "235--242",
    year = "2017"
}

@article{Pich:2023kim,
    author = "Pich, Antonio and Solomonidi, Eleftheria and Vale Silva, Luiz",
    title = "{Final-state interactions in the CP asymmetries of charm-meson two-body decays}",
    eprint = "2305.11951",
    archivePrefix = "arXiv",
    primaryClass = "hep-ph",
    doi = "10.1103/PhysRevD.108.036026",
    journal = "Phys. Rev. D",
    volume = "108",
    number = "3",
    pages = "036026",
    year = "2023"
}

@misc{FCC,
  author       = {W. Weber and K. Kr{\"o}ninger and R. Madar and S. Monteil},
  title        = {}, 
  howpublished = {{\it CP-violation in $D$-decays}, talk presented at {\it Flavours at FCC Workshop}, CERN, 19--21 November 2025.\\
                  \url{https://indico.cern.ch/event/1588013/}},
}

@misc{Punzi,
  author       = {G. Punzi},
  howpublished = {{\it Mixing and CP violation in charm decays at LHCb}, 
                  talk presented at {\it EPS-HEP Conference}, Marseille, 
                  6--11 July 2025.\\
\url{https://indico.in2p3.fr/event/33627/program}},

}

@article{LHCb:2024yxi,
    collaboration = "LHCb",
    title = "{Simultaneous determination of the CKM angle $\gamma$ and parameters related to mixing and $C\!P$ violation in the charm sector}",
    reportNumber = "LHCb-CONF-2024-004, CERN-LHCb-CONF-2024-004",
    doi = "10.17181/CERN.8CUC.W3FT",
    year = "2024"
}

@article{Bause:2022jes,
    author = {Bause, Rigo and Gisbert, Hector and Hiller, Gudrun and H{\"o}hne, Tim and Litim, Daniel F. and Steudtner, Tom},
    title = "{U-spin-CP anomaly in charm}",
    eprint = "2210.16330",
    archivePrefix = "arXiv",
    primaryClass = "hep-ph",
    reportNumber = "DO-TH 22/25",
    doi = "10.1103/PhysRevD.108.035005",
    journal = "Phys. Rev. D",
    volume = "108",
    number = "3",
    pages = "035005",
    year = "2023"
}

@article{Sinha:2025cuo,
    author = "Sinha, Rahul and Browder, Thomas E. and Deshpande, N. G. and Sahoo, Dibyakrupa and Sinha, Nita",
    title = "{Implications of the evidence for direct $\mathbf{CP}$ violation in $D\to \pi^+\pi^-$ decays}",
    eprint = "2505.24338",
    archivePrefix = "arXiv",
    primaryClass = "hep-ph",
    month = "5",
    year = "2025"
}

@article{Gavrilova:2023fzy,
    author = "Gavrilova, Margarita and Grossman, Yuval and Schacht, Stefan",
    title = "{Determination of the D{\textrightarrow}{\ensuremath{\pi}}{\ensuremath{\pi}} ratio of penguin over tree diagrams}",
    eprint = "2312.10140",
    archivePrefix = "arXiv",
    primaryClass = "hep-ph",
    doi = "10.1103/PhysRevD.109.033011",
    journal = "Phys. Rev. D",
    volume = "109",
    number = "3",
    pages = "033011",
    year = "2024"
}

@article{Iguro:2024uuw,
    author = {Iguro, Syuhei and Nierste, Ulrich and Overduin, Emil and Sch{\"u}{\ss}ler, Maurice},
    title = "{$SU(3)_F$ sum rules for CP asymmetries of $D_{(s)}$ decays}",
    eprint = "2408.03227",
    archivePrefix = "arXiv",
    primaryClass = "hep-ph",
    reportNumber = "KEK-TH-2643, TTP24-029, KA-TP-15-2024",
    doi = "10.1103/PhysRevD.111.035023",
    journal = "Phys. Rev. D",
    volume = "111",
    number = "3",
    pages = "035023",
    year = "2025"
}

@article{Schacht:2022kuj,
    author = "Schacht, Stefan",
    title = "{A U-spin anomaly in charm CP violation}",
    eprint = "2207.08539",
    archivePrefix = "arXiv",
    primaryClass = "hep-ph",
    doi = "10.1007/JHEP03(2023)205",
    journal = "JHEP",
    volume = "03",
    pages = "205",
    year = "2023"
}

@article{Fleischer:1992gp,
    author = "Fleischer, Robert",
    title = "{CP violating asymmetries in penguin induced B meson decays beyond the leading logarithmic approximation}",
    reportNumber = "TUM-T31-26-92",
    doi = "10.1007/BF01557708",
    journal = "Z. Phys. C",
    volume = "58",
    pages = "483--498",
    year = "1993"
}

@article{Grossman:2011zk,
    author = "Grossman, Yuval and Nir, Yosef",
    title = "{CP Violation in $\tau^\pm \to \pi^\pm K_S\nu$ and $D^\pm \to \pi^\pm K_S$: The Importance of $K_S - K_L$ Interference}",
    eprint = "1110.3790",
    archivePrefix = "arXiv",
    primaryClass = "hep-ph",
    doi = "10.1007/JHEP04(2012)002",
    journal = "JHEP",
    volume = "04",
    pages = "002",
    year = "2012"
}

@article{Dery:2019ysp,
    author = "Dery, Avital and Nir, Yosef",
    title = "{Implications of the LHCb discovery of CP violation in charm decays}",
    eprint = "1909.11242",
    archivePrefix = "arXiv",
    primaryClass = "hep-ph",
    doi = "10.1007/JHEP12(2019)104",
    journal = "JHEP",
    volume = "12",
    pages = "104",
    year = "2019"
}

@article{Belle:2024vho,
    author = "Adachi, I. and others",
    collaboration = "Belle, Belle-II",
    title = "{Measurement of the time-integrated CP asymmetry in $D^0\to K_{\rm S}^0K_{\rm S}^0$ decays using Belle and Belle II data}",
    eprint = "2411.00306",
    archivePrefix = "arXiv",
    primaryClass = "hep-ex",
    reportNumber = "KEK Preprint 2024-26, Belle II Preprint 2024-026",
    doi = "10.1103/PhysRevD.111.012015",
    journal = "Phys. Rev. D",
    volume = "111",
    number = "1",
    pages = "012015",
    year = "2025"
}

@article{CMS:2024hsv,
    author = "Hayrapetyan, Aram and others",
    collaboration = "CMS",
    title = "{Search for $CP$ violation in D$^0$$\to$ K$^0_\mathrm{S}$K$^0_\mathrm{S}$ decays in proton-proton collisions at $\sqrt{s}$ = 13 TeV}",
    eprint = "2405.11606",
    archivePrefix = "arXiv",
    primaryClass = "hep-ex",
    reportNumber = "CMS-BPH-23-005, CERN-EP-2024-120",
    doi = "10.1140/epjc/s10052-024-13244-0",
    journal = "Eur. Phys. J. C",
    volume = "84",
    number = "12",
    pages = "1264",
    year = "2024"
}

@article{Grossman:2019xcj,
    author = "Grossman, Yuval and Schacht, Stefan",
    title = "{The emergence of the $\Delta U=0$ rule in charm physics}",
    eprint = "1903.10952",
    archivePrefix = "arXiv",
    primaryClass = "hep-ph",
    doi = "10.1007/JHEP07(2019)020",
    journal = "JHEP",
    volume = "07",
    pages = "020",
    year = "2019"
}

@article{Schacht:2021jaz,
    author = "Schacht, Stefan and Soni, Amarjit",
    title = "{Enhancement of charm CP violation due to nearby resonances}",
    eprint = "2110.07619",
    archivePrefix = "arXiv",
    primaryClass = "hep-ph",
    doi = "10.1016/j.physletb.2021.136855",
    journal = "Phys. Lett. B",
    volume = "825",
    pages = "136855",
    year = "2022"
}

@article{Chala:2019fdb,
    author = "Chala, Mikael and Lenz, Alexander and Rusov, Aleksey V. and Scholtz, Jakub",
    title = "{$\Delta A_{CP}$ within the Standard Model and beyond}",
    eprint = "1903.10490",
    archivePrefix = "arXiv",
    primaryClass = "hep-ph",
    reportNumber = "IPPP/19/25",
    doi = "10.1007/JHEP07(2019)161",
    journal = "JHEP",
    volume = "07",
    pages = "161",
    year = "2019"
}

@article{Bediaga:2022sxw,
    author = "Bediaga, Ignacio and Frederico, Tobias and Magalh{\~a}es, Patricia C.",
    title = "{Enhanced Charm CP Asymmetries from Final State Interactions}",
    eprint = "2203.04056",
    archivePrefix = "arXiv",
    primaryClass = "hep-ph",
    doi = "10.1103/PhysRevLett.131.051802",
    journal = "Phys. Rev. Lett.",
    volume = "131",
    number = "5",
    pages = "051802",
    year = "2023"
}

@article{Cheng:2019ggx,
    author = "Cheng, Hai-Yang and Chiang, Cheng-Wei",
    title = "{Revisiting CP violation in $D\to P\!P$ and $V\!P$ decays}",
    eprint = "1909.03063",
    archivePrefix = "arXiv",
    primaryClass = "hep-ph",
    doi = "10.1103/PhysRevD.100.093002",
    journal = "Phys. Rev. D",
    volume = "100",
    number = "9",
    pages = "093002",
    year = "2019"
}

@article{Nierste:2015zra,
    author = "Nierste, Ulrich and Schacht, Stefan",
    title = "{CP Violation in $D^0\rightarrow K_SK_S$}",
    eprint = "1508.00074",
    archivePrefix = "arXiv",
    primaryClass = "hep-ph",
    reportNumber = "TTP15-027",
    doi = "10.1103/PhysRevD.92.054036",
    journal = "Phys. Rev. D",
    volume = "92",
    number = "5",
    pages = "054036",
    year = "2015"
}

@article{Friday:2025gpj,
    author = "Friday, David and Gersabeck, Evelina and Lenz, Alexander and Piscopo, Maria Laura",
    title = "{Charm physics}",
    eprint = "2506.15584",
    archivePrefix = "arXiv",
    primaryClass = "hep-ph",
    reportNumber = "Nikhef 2025-009, SI-HEP-2025-14, P3H-25-040",
    month = "6",
    year = "2025"
}

@article{LHCb:2025ezf,
    author = "Aaij, Roel and others",
    collaboration = "LHCb",
    title = "{Measurement of $C\!P$ asymmetry in $D^0 \to K^0_{\rm S} K^0_{\rm S}$ decays with the LHCb Upgrade I detector}",
    eprint = "2510.14732",
    archivePrefix = "arXiv",
    primaryClass = "hep-ex",
    reportNumber = "LHCb-PAPER-2025-036, CERN-EP-2025-221",
    month = "10",
    year = "2025"
}

@article{Buras:1994ij,
    author = "Buras, Andrzej J.",
    title = "{QCD factors $a_1$ and $a_2$ beyond leading logarithms versus factorization in nonleptonic heavy meson decays}",
    eprint = "hep-ph/9409309",
    archivePrefix = "arXiv",
    reportNumber = "MPI-PHT-94-60, TUM-T31-75-94",
    doi = "10.1016/0550-3213(94)00482-T",
    journal = "Nucl. Phys. B",
    volume = "434",
    pages = "606--618",
    year = "1995"
}

\end{document}